\documentclass[acmsmall]{acmart}

\AtBeginDocument{%
  }

\setcopyright{acmlicensed}
\copyrightyear{2018}
\acmYear{2018}
\acmDOI{XXXXXXX.XXXXXXX}

\begin{document}

\title{A Performance Analysis of BFT Consensus for Blockchains}

\author{J.D. Chan}
\author{Y.C. Tay}
\email{dcstayyc@nus.edu.sg}
\orcid{0000-0002-6280-2469}
\author{Brian R.Z. Yen}
\affiliation{%
	\institution{National University of Singapore}
  \country{Singapore}
}

\begin{abstract}
	Distributed ledgers are common in the industry.
Some of them can use blockchains as their underlying infrastructure.
A blockchain requires participants to agree on its contents.
This can be achieved via a consensus protocol,
and several BFT (Byzantine Fault Tolerant) protocols have been proposed
for this purpose.
How do these protocols differ in performance?
And how is this difference affected by the communication network?
Moreover, such a protocol would need a timer to ensure progress,
but how should the timer be set?

This paper presents an analytical model to address these and related issues
in the case of crash faults.
Specifically, it focuses on two consensus protocols (Istanbul BFT and HotStuff)
and two network topologies (Folded-Clos and Dragonfly).
The model provides closed-form expressions for analyzing how the timer value
and number of participants, faults and switches affect
the consensus time.
The formulas and analyses are validated with simulations.
The conclusion offers some tips for analytical modeling of such protocols.
\end{abstract}

\ccsdesc[500]{Networks~Network protocols}
\ccsdesc[500]{Networks~Application layer protocols}
\ccsdesc[500]{Computing methodologies~Modeling methodologies}
\ccsdesc[500]{Computing methodologies~Distributed algorithms}

\keywords{Byzantine faults, consensus protocols, blockchain,
performance model}

\maketitle

\def \roundchange {{\tt ROUND-CHANGE}}
\def \preprepare {{\tt PRE-PREPARE}}
\def \prepare {{\tt PREPARE}}
\def \commit {{\tt COMMIT}}
\def \newview {{\tt NEW-VIEW}}
\def \precommit {{\tt PRECOMMIT}}
\def \decide {{\tt DECIDE}}
\def \btl {{\tau_0}}
\def \ns {{\nu_s}}
\def \nue {{\nu_e}}
\def \nl {{\nu_\ell}}
\def \nn {{\nu_{12}}}
\def \nd {{\nu_d}}
\def \ng {{\nu_g}}
\def \nT {{n_\mathcal{T}}}
\def \nF {{n_\mathcal{F}}}
\def \nD {{n_\mathcal{D}}}
\def \mH {{m^\mathcal{H}}}
\def \mI {{m^\mathcal{I}}}
\def \mIT {{m^\mathcal{I}_\mathcal{T}}}
\def \mIF {{m^\mathcal{I}_\mathcal{F}}}
\def \mID {{m^\mathcal{I}_\mathcal{D}}}
\def \nIT {{n^\mathcal{I}_\mathcal{T}}}
\def \nIF {{n^\mathcal{I}_\mathcal{F}}}
\def \nID {{n^\mathcal{I}_\mathcal{D}}}
\def \hT {{h_\mathcal{T}}}
\def \calT {\mathcal{T}}
\def \calP {\mathcal{P}}
\def \calH {\mathcal{H}}
\def \calI {\mathcal{I}}
\def \calC {\mathcal{C}}
\def \calD {\mathcal{D}}
\def \calF {\mathcal{F}}
\def \calE {\mathcal{E}}
\def \calN {\mathcal{N}}
\def \TPT  {{T^\mathcal{P}_\mathcal{T}}}
\def \TPC  {{T^\mathcal{P}_\mathcal{C}}}
\def \THT  {{T^\mathcal{H}_\mathcal{T}}}
\def \THC  {{T^\mathcal{H}_\mathcal{C}}}
\def \THD  {{T^\mathcal{H}_\mathcal{D}}}
\def \THF  {{T^\mathcal{H}_\mathcal{F}}}
\def \TIT  {{T^\mathcal{I}_\mathcal{T}}}
\def \TIC  {{T^\mathcal{I}_\mathcal{C}}}
\def \TID  {{T^\mathcal{I}_\mathcal{D}}}
\def \TIF  {{T^\mathcal{I}_\mathcal{F}}}
\def \TPTTT{{({T^\mathcal{P}_{\mathcal{T}})_3}}}
\def \TPCCC{{({T^\mathcal{P}_{\mathcal{C}})_3}}}
\def \THTTT{{({T^\mathcal{H}_{\mathcal{T}})_3}}}
\def \THCCC{{({T^\mathcal{H}_{\mathcal{C}})_3}}}
\def \TICCC{{({T^\mathcal{I}_{\mathcal{C}})_3}}}
\def \TITTT{{({T^\mathcal{I}_{\mathcal{T}})_3}}}
\def \rPv  {{r^\mathcal{P}_v}}
\def \rPs  {{r^\mathcal{P}_s}}
\def \rHv  {{r^\mathcal{H}_v}}
\def \rHs  {{r^\mathcal{H}_s}}
\def \rIv  {{r^\mathcal{I}_v}}
\def \rIs  {{r^\mathcal{I}_s}} 
\def \Pr {{\rm Pr}}
\def \XHv {{X_v^\mathcal{H}}}
\def \XHs {{X_s^\mathcal{H}}}

\section{Introduction}
\label{sec:intro}

There is considerable current interest in the use of blockchains
to implement distributed ledgers for finance, health care, energy, logistics, 
etc.~\cite{BlockchainPerfSurvey.IEEEAccess2020}. 
This requires continual consensus among interested parties in the system.
Research on consensus protocols is mostly split between proof-based consensus 
(like Bitcoin) and vote-based consensus~\cite{BlockchainConsensus.CSurveys2023}.

A vote-based protocol for enforcing consensus must guard against errant
(possibly malicious) validator behavior.
It is said to be {\it Byzantine Fault Tolerant} (BFT)
if it can achieve consensus despite arbitrary misbehavior.
There is a large variety of BFT protocols.
They differ in terms of failure models, delay assumptions,
cryptographic requirements, etc.
Much of the impetus in developing new protocols lies in 
pushing their performance.

The performance of these protocols are measured via complexity analysis
(e.g. number of messages or rounds), in empirical or simulation experiments,
or with analytical models.
Despite decades of such analysis of BFT consensus,
much remains unknown regarding their behavior.

For example, to ensure the protocol progresses despite failures,
a common technique lies in setting a {\it timer};
when it expires, the protocol assumes there is a fault,
and restarts the consensus process.
This timer has some initial value $\btl$,
but the value is usually increased upon timer expiry.
How should $\btl$ be set?

\begin{figure}
        \centering
        \begin{tabular}{ccc}
\includegraphics[width=0.30\textwidth]{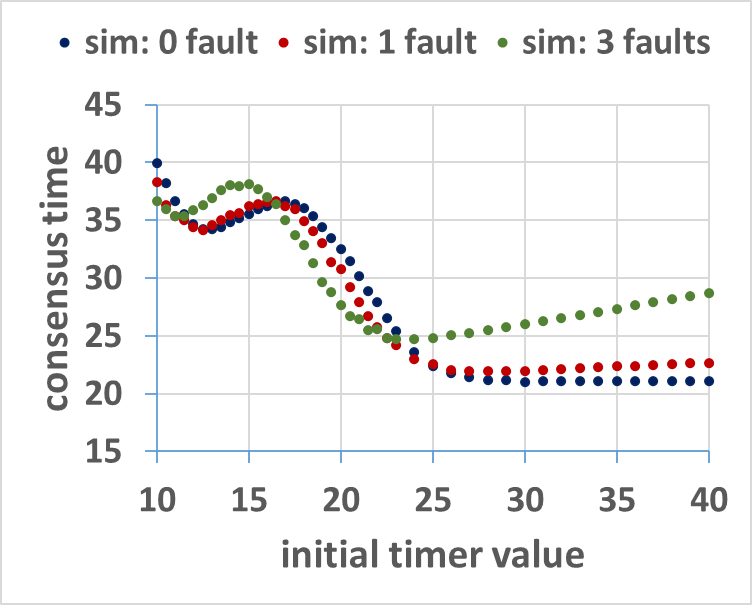} &
\includegraphics[width=0.30\textwidth]{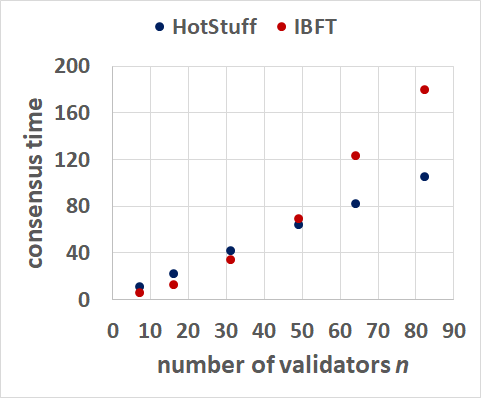}&
\includegraphics[width=0.30\textwidth]{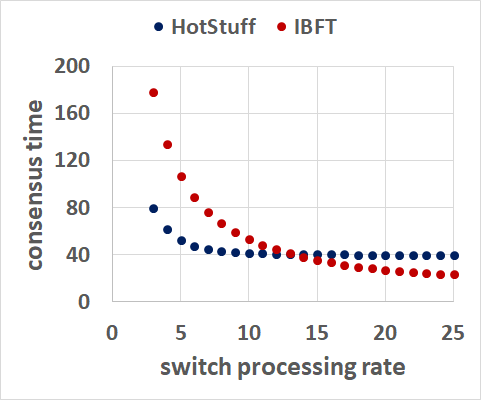} \\
(a) non-monotonic behavior  
& (b) crossover for $n$
& (c) crossover  for switch rate \\
        \end{tabular}
\caption{Simulated consensus time for 
	(a) HotStuff with varying number of faults ($n=16$, clique),
	(b) HotStuff vs IBFT on Dragonfly ($\nd=5, \rPs=9$) 
	 and (c) HotStuff vs IBFT on Dragonfly ($\nd=3$, $n=31$).
	 The unit of time on the vertical axis follows that for 
	 message processing time at a validator (see Sec.~\ref{sec:simulator}).}
	\label{fig:IntroSim}
\end{figure}

For guidance, we simulated HotStuff~\cite{HotStuff.PODC2019},
a well-known BFT protocol for blockchains,
and measured the time $T$ to gather consensus for a block.
Fig.~\ref{fig:IntroSim}(a) shows how $T$ varies with $\btl$
and the number of faults.
How can one analytically describe this interesting, non-monotonic behavior?

\begin{figure}
\centerline{
	\includegraphics[width=5.9cm]{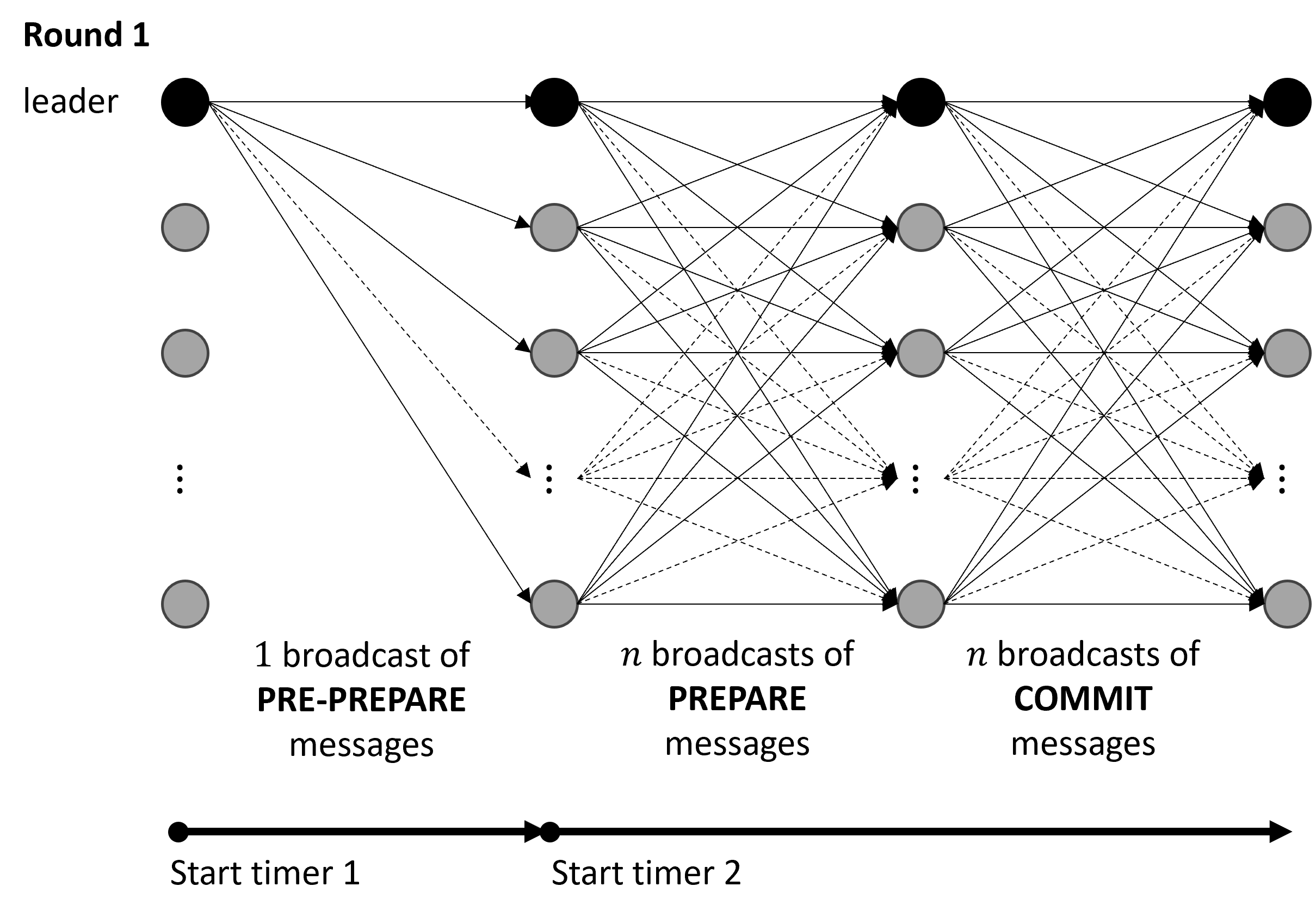}
	\qquad
	\includegraphics[width=7.6cm]{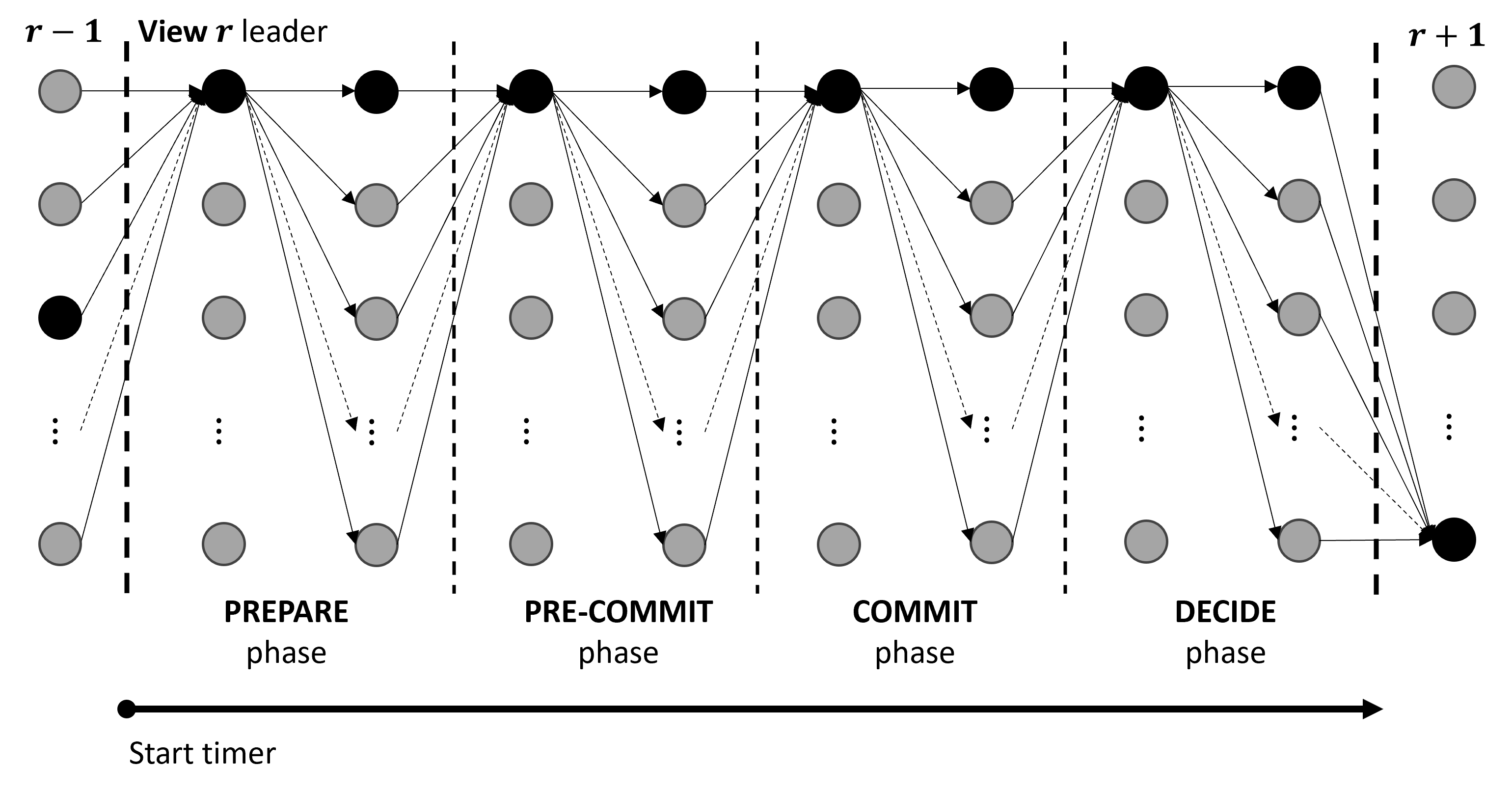}
}
	\centerline{\hfill (a) IBFT \hglue 6cm (b) HotStuff \hfill}
\caption{IBFT and HotStuff have very different patterns for message exchange.}
	\label{fig:MsgPattern}
\end{figure}

In the simulations for Fig.~\ref{fig:IntroSim}(a),
we factor out the impact of the network connecting the nodes
by using 0 message delay between any two nodes.
What if the network delays are not negligible?
How might two topologies differ in their impact on $T$?

A network can affect $T$ through congestion delays.
We therefore expect that the same topology can have different impact
on two protocols, if their message patterns differ.
For example, consider the pattern in Fig.~\ref{fig:MsgPattern}(a)
for IBFT (Instanbul BFT) protocol~\cite{IBFT.arXiv2020},
which is very different from the one in Fig.~\ref{fig:MsgPattern}(b)
for HotStuff.
One may expect that, for the same topology,
congestion delays can have greater impact on IBFT than HotStuff.

However, simulation results in Fig.~\ref{fig:IntroSim}(b) show that,
for the Dragonfly~\cite{Dragonfly.SC2019},
IBFT can be faster or slower than HotStuff, 
depending on the number of participants.
Fig.~\ref{fig:IntroSim}(c) shows a similar slower/faster crossover  
depending on how fast a switch or router can process a message. 
How do the protocol and topology parameters determine the crossover point
in their consensus time?

This paper presents three contributions:
(1) An analysis of how BFT consensus time depends on the number of validators
and faults, timer value and network topology.
(2) A comparison of two protocols' performance, and two topologies' impact. 
(3) An approach to analytic modeling of consensus protocols.  

Below, we begin with a review of related work in Sec.~\ref{sec:related}. 
Sec.~\ref{sec:prelim} introduces necessary definitions and notation,
and describes two protocols (IBFT and HotStuff)
and two topologies (Dragonfly and Folded-Clos)
that we use for analyzing consensus time;
it also presents the approach to the performance analysis and its validation.
A Byzantine validator can cause arbitrary performance 
degradation~\cite{Clement.NSDI2009}, so we only consider crash faults.
The performance model is presented in Sec.~\ref{sec:clique}
for cliques and Sec.~\ref{sec:nonclique} for non-clique topologies.
Sec.~\ref{sec:conclusions} highlights some conclusions about the consensus time 
for HotStuff and IBFT, and the modeling approach.

\section{Related Work}
\label{sec:related}

BFT protocols were introduced 40-odd years ago~\cite{BFT.JACM1980}.
The problem formulation helped to kick-start research on distributed computing,
and many variations followed thereafter.
The PBFT (Practical BFT) protocol~\cite{PBFT.OSDI1999} revived interest
in the problem, and this was further boosted by the application of BFT
consensus to blockchains~\cite{BlockchainConsensus.CSurveys2023}.
IBFT and HotStuff are examples of BFT protocols that are specifically
designed for blockchains~\cite{IBFT.arXiv2020, LibraBFT.NFM2022}.

Pre-PBFT, the design of BFT protocols was mostly driven by changes in 
assumptions (e.g. network synchrony or failure detectors) 
and improvements in complexity (e.g. number of rounds or bits).
However, the performance of a protocol goes beyond algorithmic complexity.
Experimental measurements show that consensus latency and throughput
can be significantly affected by message size, verification algorithm, 
network topology, attack model, etc.~\cite{700BFT.ToCS2015,
Algorand.SOSP2017, RCC.ICDE2021, MirBFT.JSR2022}.

In any experimental study, measurements are only done for some particular
choice and combination of values for the variables (e.g. number of switches)
or parameters (e.g. timer value), so the observations may not be conclusive.
For example, one protocol may be faster than another in some observed 
part of the parameter space, but not in some other, unobserved part.

One could get a global view of the parameter space by using an analytical model
of the protocol.  
Unfortunately, we could not find an adequate model in the literature.
For example, Ricci et al. model the delays experienced by clients who submit
transactions that make up the blocks, but they do not model the
consensus protocol~\cite{BlockchainQDelay.SOCCA2018},
and Ni et al.'s model for blockchains in a 6G context does not include the 
timer~\cite{6GBlockchain.IoTJ2022}.

In terms of modeling technique, an obvious choice would be 
Markov chains~\cite{PBFTconsensus.P2Pnetapp2022},
but such models rarely yield closed-form expressions that provide a global view;
one ends up solving them numerically for some choice of parameter values so,
again, we can only see parts of the parameter space.
The same is true of queueing and Petri net 
models~\cite{BlockchainQing.TPDS2021, HyperledgerPetri.CompComm2020}.

In contrast, we present a modeling approach that is based on 
crude approximations and elementary probabilistic arguments.
It also uses bottleneck analysis, which is simple, yet effective,
as demonstrated by the impactful Roofline model~\cite{Roofline.CACM2009}.
This approach leads to closed-form expressions that facilitate
performance analysis of consensus protocols on realistic network topologies.

\section{Preliminaries}
\label{sec:prelim}

We are interested in how the performance of BFT consensus depends on 
the protocol and the network topology.
Below, Sec.~\ref{sec:2protocols} first describes two BFT protocols,
and Sec.~\ref{sec:2topologies} follows with two topologies.
Sec.~\ref{sec:key.eqn} then states the key equation for estimating
consensus time, taking faults into account.
Sec.~\ref{sec:simulator} describes the simulator that we use
to validate the performance model.
The reader can skip forward to 
Sec.~\ref{sec:clique} and Sec.~\ref{sec:nonclique},
and refer to this section (or Table~\ref{tab:term.notation} in the Appendix)
for definitions and notation when needed.

\subsection{BFT Protocols}
\label{sec:2protocols}

We want to know how consensus time depends on the number of validators $n$.
We therefore consider a {\it closed system}, where $n$ is fixed;
there are standard techniques for using a closed system as a submodel for an 
{\it open system}~\cite{TayBook}, where $n$ varies as validators come and go.
Let $n_f$ denote the number of faulty validators,
and $n_w=n-n_f$ the number of nonfaulty or {\it working} validators.

The correctness of BFT protocols depends critically on their assumptions
regarding message delays~\cite{HoneyBadger.2016}.
We therefore pick two protocols with the same assumption, 
namely partial synchrony~\cite{PartialSync.JACM1988}.  
IBFT and HotStuff are chosen because they have
canonical but very different patterns for message exchange
(see Fig.~\ref{fig:MsgPattern}).  
 
These 2 protocols specify a bound $f\ge n_f$,
and are provably correct if $n\ge 3f+1$.
A {\it quorum} consists of $2f+1$ valid votes for a consensus.  

\subsubsection{Terminology and Notation (Protocols)}
\phantom{.}\\ 
Each protocol is run by $n$ {\it validators} or {\it nodes}.
A {\it block} is a collection of data (client or ledger information, etc.).
A {\it blockchain} is a sequence of blocks.
Each validator adds blocks to its own copy of the blockchain,
and the essence of the consensus lies in ensuring all copies are the same.

The $j$th {\it consensus instance} is the state in which the $(j-1)$th block
is in the blockchain and the validators are working out the consensus
for adding the $j$th block.

A {\it round} (in IBFT) or {\it view} (in HotStuff) is a single iteration
of the protocol for a consensus instance.
Each successful completion of 1 round implies consensus.
Each round aims to garner validator consensus 
for adding a block to the blockchain.
Unless there are timeouts, 
a fault-free run of the protocol would lead to consensus in 1 round.

Each validator has a {\it timer};
if it does not reach consensus upon timer expiry,
it will initiate a {\it round change}.
Faults and timeouts can cause multiple round changes 
before consensus is reached.
It is possible for multiple validators to undergo round change
while others have already reached consensus.
Crucially, a {\it full round change} refers to the case where all validators
undergo round change for the same round.
This can happen if the leader crashes, or validators time out.
The initial value for a timer is $\btl$.
The value is doubled everytime the timer expires (i.e. exponential backoff).

We now describe the two protocols,
focusing on the message exchange and omitting details for verification,
encryption, etc.

\subsubsection{IBFT Consensus Protocol}
\label{sec:IBFTconsensus}
\phantom{.}\\
IBFT is an adaptation of PBFT for the Quorum blockchain~\cite{IBFT.arXiv2020}.
A round has a single validator that is a {\it leader}; 
this role is rotated among validators.
Conceptually, a round has 3 phases (see Fig.~\ref{fig:MsgPattern}(a)):

\noindent (\preprepare)\quad
\hangindent=\parindent
In the \preprepare\ phase, the leader starts its first timer
(see Fig.~\ref{fig:MsgPattern}(a))
and broadcasts a \preprepare\ to all validators.
Upon receiving a \preprepare,
a validator starts its second timer,
transitions to the \prepare\ phase,
and broadcasts a \prepare\ to all validators.

\noindent (\prepare)\quad
\hangindent=\parindent
While in the \prepare\ phase, a validator that receives a quorum of $2f+1$
\prepare\ messages transitions into the \commit\ phase
and broadcasts a \commit\ message.

\noindent (\commit)\quad
\hangindent=\parindent
A validator that receives a quorum of \commit\ messages 
considers that as indicating consensus;
adds the block to its blockchain, starts its first timer
and returns to the \preprepare\ phase for the next consensus instance.

\noindent
If a validator's timer expires before consensus is reached,
it initiates a separate round change protocol 
by broadcasting a \roundchange\ message:
upon receiving a quorum of these messages,
the new leader broadcasts a \preprepare\ message and the protocol resumes.

If the (pre-determined) leader for the next round 
receives a quorum of \roundchange\ messages,
it abandons the current round, doubles its timer duration,
enters a \preprepare\ phase, and starts a fresh round for the block.

\subsubsection{HotStuff Consensus Protocol}
\label{sec:HSconsensus}
\phantom{.}\\
HotStuff is used as a basis for blockchain protocols
like LibraBFT for the Diem blockchain~\cite{LibraBFT.NFM2022}. 
A {\it view} in HotStuff is like a round in IBFT,
so we use the two terms interchangeably.
As in IBFT, each view has a {\it timer},
and a {\it leader} that is also a validator.
A leader is randomly chosen as views change.
A view has 4 phases (see Fig.~\ref{fig:MsgPattern}(b)):

\noindent (\prepare)\quad
\hangindent=\parindent
In the \prepare\ phase, a validator starts its timer 
and sends a \newview\ message to the leader.
Upon receiving 
$2f+1$ \newview\ messages,
the leader broadcasts a \prepare\ message.

\noindent (\precommit)\quad
\hangindent=\parindent
Upon receiving a \prepare\ message,
a validator transitions to a \precommit\ phase 
and sends a \prepare\ message to the leader.
When the leader receives a quorum of \prepare\ messages,
it broadcasts a \precommit\ message.

\noindent (\commit)\quad
\hangindent=\parindent
A validator that receives a \precommit\ message 
transitions into a \commit\ phase 
and sends a \precommit\ message to the leader.
When the leader receives a quorum of \precommit\ messages,
it broadcasts a \commit\ message.

\noindent (\decide)\quad
\hangindent=\parindent
Upon receiving the \commit\ message,
a validator transitions into a \decide\ phase
and sends a \commit\ message to the leader.
When the leader receives a quorum of \commit\ messages,
it broadcasts a \decide\ message.
A validator that receives a \decide\ message 
considers that as indicating consensus,
adds the block to its blockchain,
resets its timer and returns to the \prepare\ phase for the next view.

\noindent
Unlike IBFT, 
a timer expiry does not initiate a separate protocol for view change.
Instead, the validator with the expired timer doubles its timer duration,
transitions into the \prepare\ phase for the next view,
and sends a \newview\ message (for the same block) 
to the leader for the next view.

\subsection{Network Topology}
\label{sec:2topologies}

Most BFT protocol designs are independent of the network topology
connecting the nodes (where the validators are).
In effect, the nodes are directly connected to each other
via a {\it clique}.
We consider this a virtual topology, as it is not scalable
(consider the quadratic node-node links).

We are interested in how the underlying physical topology affects 
protocol performance.  
We therefore only consider those that can be unambiguously parameterized.
Wide-area networks do not fit this requirement well.
We therefore restrict our attention to BFT protocols 
that run within a single datacenter (e.g. enterprise blockchains).
There are many possible datacenter network topologies.
We pick two that have contrasting designs:
Folded Clos~\cite{MeshedTree.FIR2023} and Dragonfly~\cite{Dragonfly.SC2019}.

\subsubsection{Terminonology and Notation (Topology)}
\phantom{.}\\
A topology $\calT$ is a graph that connects {\it switches} 
with {\it links} as edges.
Switches help to route messages between validators.
The nodes that run validators, called {\it terminal nodes},
connect to this network via {\it edge switches}.
The edge switch that has a leader connected to it is called the 
{\it leader switch}.

The Dragonfly is a {\it direct} network consisting of only edge switches
(all have terminal nodes attached).
The Folded Clos is an {\it indirect} network that has both edge and non-edge
switches.

Let $\ns$ be the number of switches and $\nue$ the number of edge switches.
Let $h_\calT$ be the average hop distance in topology $\calT$,
i.e. the number of switches between a pair of validators.

We now describe the two topologies.

\subsubsection{Folded-Clos}
\phantom{.}\\
We consider only a 3-level {\it Folded-Clos} 
indirect network with edge switches in level~1.  
To minimize the number of topology parameters, 
we assume each level has $\nue$ switches, so $\ns=3\nue$.

Let $\nn$ be the number of links to level~2 switches from a level~1 
edge switch.
We require $\nn$ to divide $\nue$,
and the number of links to a level~3 switch from a level~2 switch is $\nue/\nn$.

Terminal nodes are connected to level~1 switches in a way that minimizes
the maximum number of validators per edge switch (i.e. even distribution).

Fig.~\ref{fig:net.topo}(a) and (b) illustrate the 
Folded-Clos$(\nue=8,\nn=4)$ and Folded-Clos$(\nue=9,\nn=3)$.

\subsubsection{Dragonfly}
\phantom{.}\\
In general, a Dragonfly is a direct network that divides switches into groups,
with links between groups and possibly multiple intra-group topologies.
Again, to minimize the number of topology parameters,
we consider the following 1-parameter variant:

Let $\nd$ be the number of switches connected by a clique in a group.
Each switch in a group has a link to a switch in another group,
so the number of groups is $\ng=\nd+1$,
and $\ns=\ng\nd=\nd(\nd+1)$.

Like for the Folded-Clos, terminal nodes are evenly distributed over the
switches.

\begin{figure}
        \centering
        \begin{tabular}{cccc}
                \includegraphics[width=0.24\textwidth]{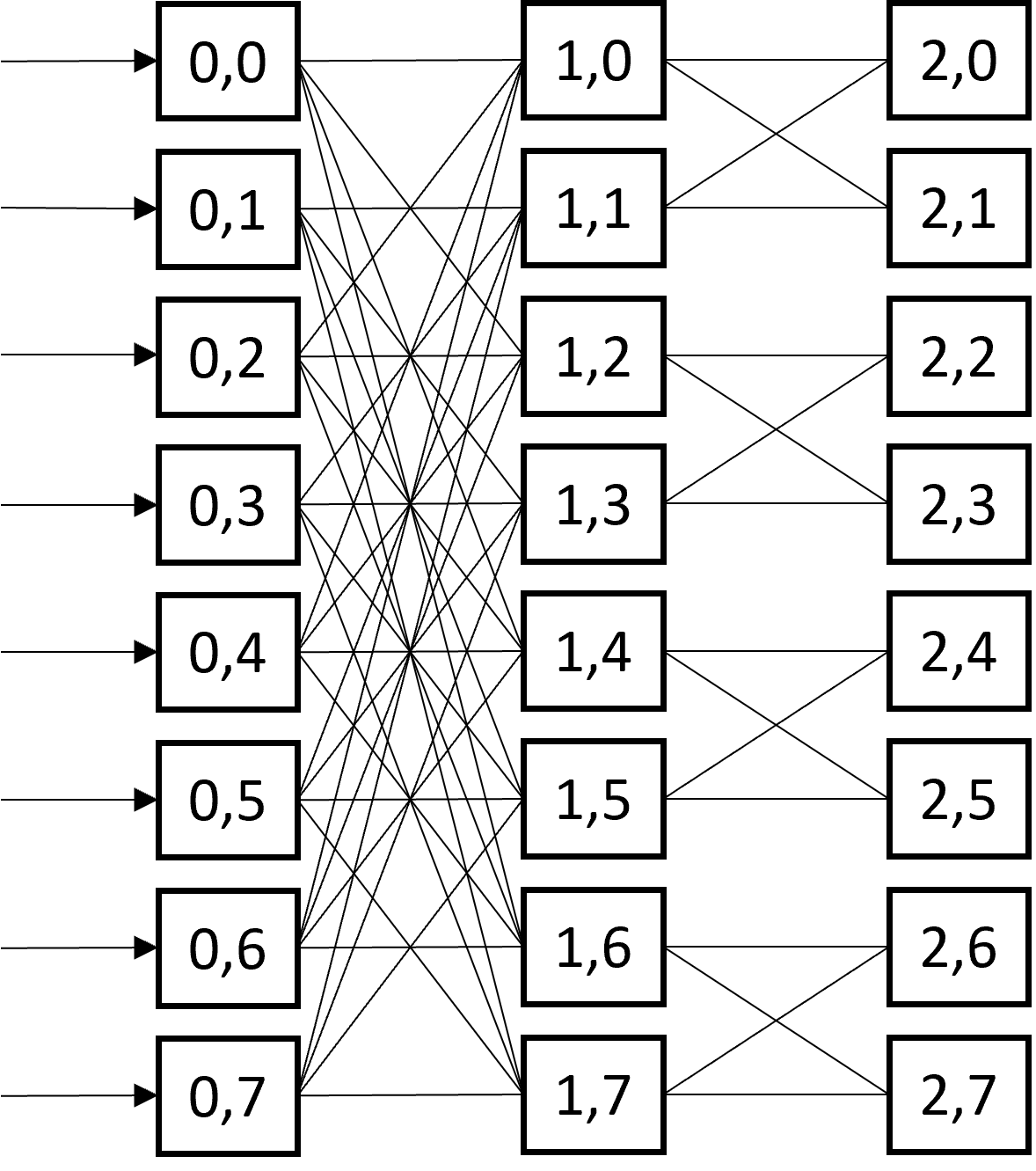} &
                \includegraphics[width=0.22\textwidth]{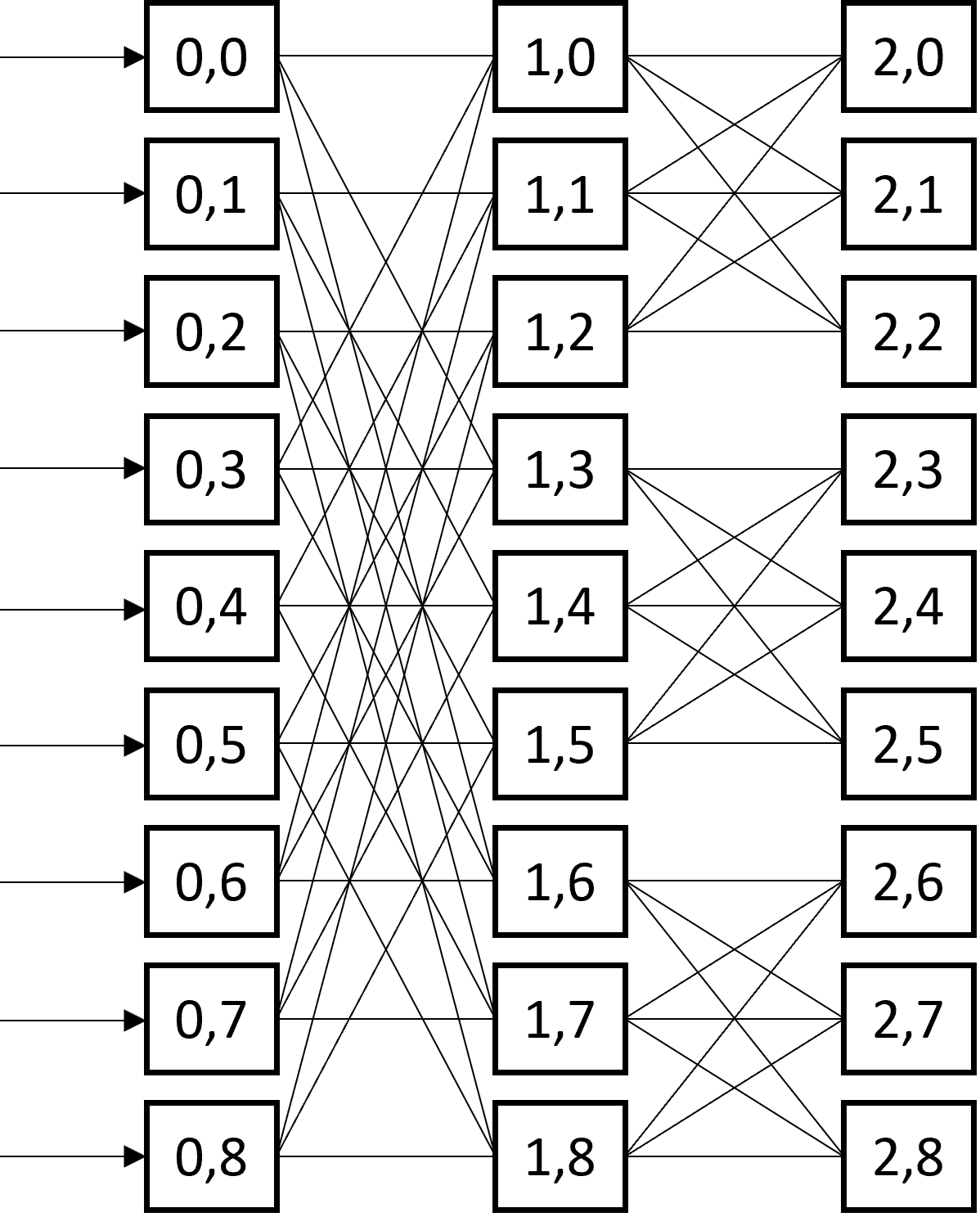} &
		\includegraphics[width=0.23\textwidth]{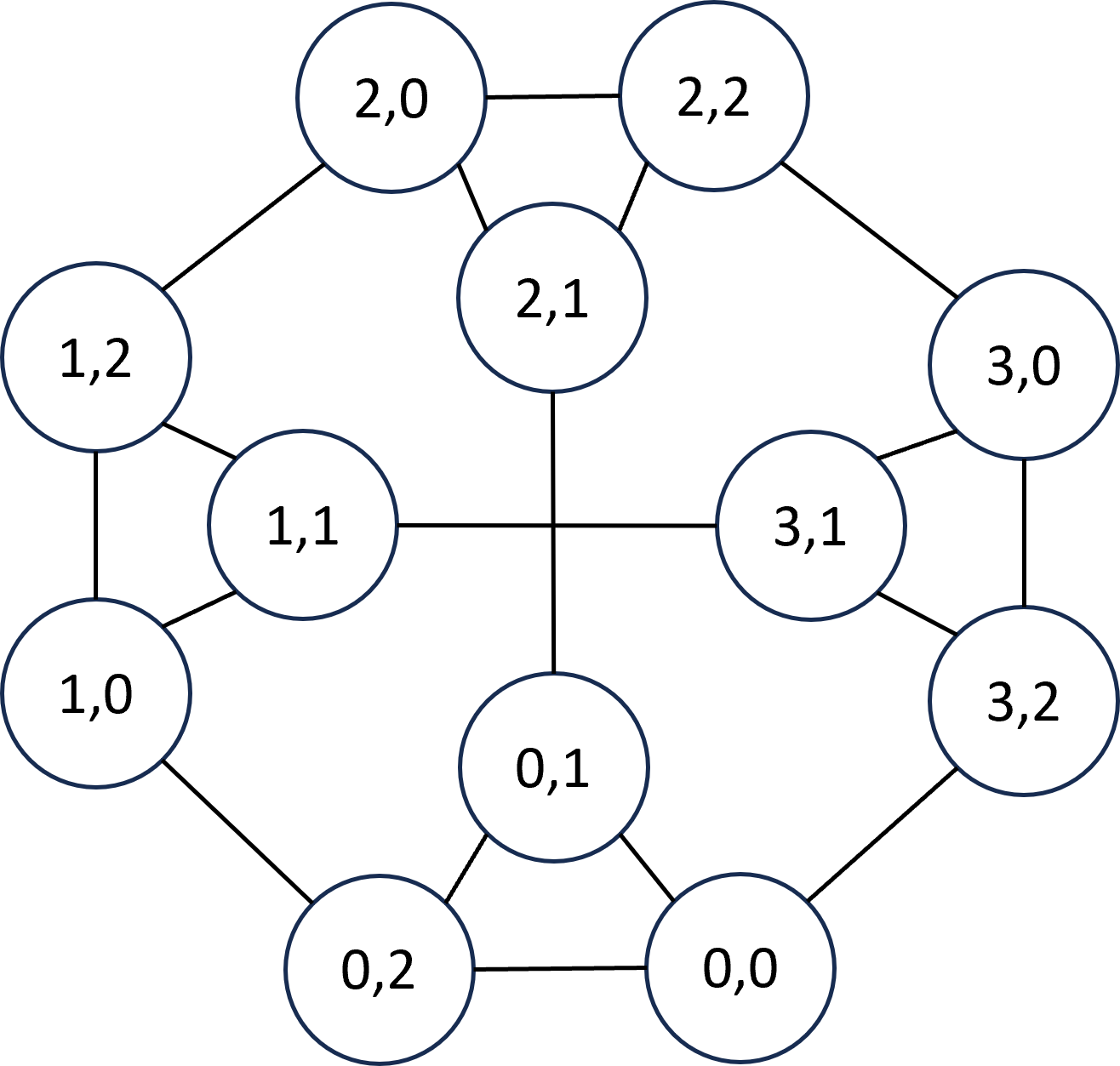} &
                \includegraphics[width=0.24\textwidth]{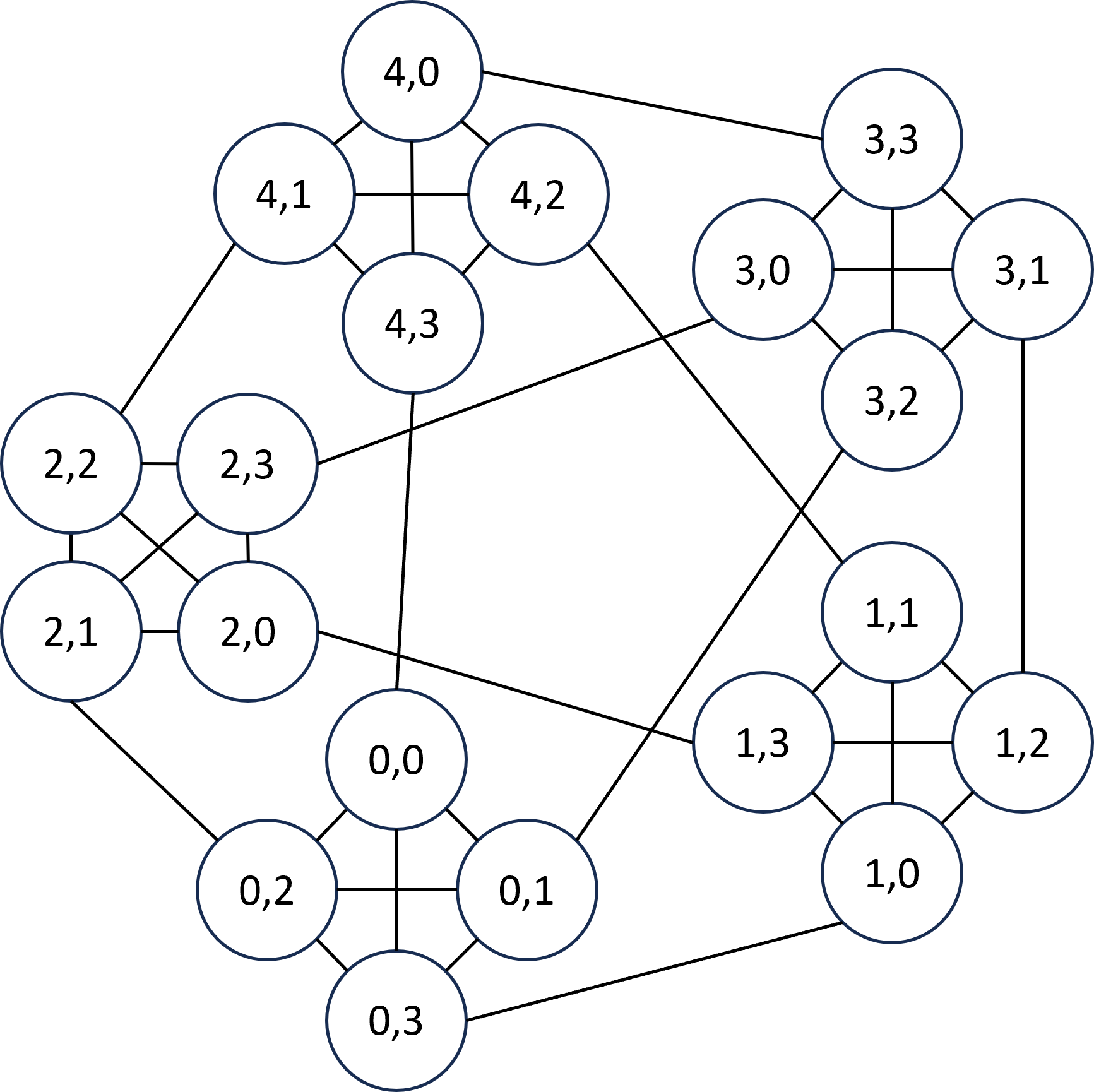} \\
  (a) Folded-Clos(8,4) &  (b) Folded-Clos(9,3)  
& (c) Dragonfly(3)         & (d) Dragonfly(4)\\
        \end{tabular}
\caption{Topology examples with parameters 
(a) $\nue=8,\nn=4$;  (b) $\nue=9,\nn=3$;  (c) $\nd=3$;  (d) $\nd=4$.} 
\label{fig:net.topo}
\end{figure}

Fig.~\ref{fig:net.topo}(c) and (d) illustrate the topology for 
Dragonfly$(\nd=3)$ and Dragonfly$(\nd=4)$.

\subsection{Performance}
\label{sec:key.eqn}

Fan et al. divide a distributed ledger architecture into 5 layers:
application, execution, data, consensus and 
network~\cite{BlockchainPerfSurvey.IEEEAccess2020}.
We will focus on the performance of the consensus layer.

A user or administrator may be interested in 
the transaction latency or throughput.
These are metrics for the data layer,
where transactions arrive in a memory pool (mempool) and are grouped into blocks
that are sent to the consensus layer~\cite{HotStuff.ApPLIED2023}.
Transaction latency therefore partly depends on the mempool 
protocol~\cite{NarwhalMempool.Eurosys2022} above the consensus layer.

In steady state, the block throughput is determined by the transaction arrival
rate in the data layer.
We therefore focus on {\it block latency} $ET$,
i.e. the average time between when a leader proposes a block 
and when consensus is reached.
The ``service rate'' $1/ET$,
together with the block throughput generated by the mempool protocol,
determine the transaction latency.
We therefore use a {\it closed} model to determine $ET$,
i.e. every consensus generates another consensus instance.

Datacenters are multi-tenanted, so any blockchain must contend with
network traffic from other workloads.
Intuitively, running $c$ blockchains concurrently should be equivalent to 
dividing switch rate by $c$,  
assuming the switches have little idle time.
In the appendix, we present an experiment that confirms this intuition
(see Fig.~\ref{fig:multichain} in the Appendix).
Therefore, the performance model will henceforth assume 
there is just 1 blockchain, with no competing network traffic.
We first introduce some notation, 
before stating the equation that underlies the performance model.

\subsubsection{Notation (Performance)}
\phantom{.}\\
Let $\TPT$ denote the consensus time for protocol $\calP$
and topology $\calT$, where $\calP=\calH$ for HotStuff and $\calI$ for IBFT, 
and $\calT=\calC$ for clique, $\calD$ for Dragonfly and $\calF$ for Folded-Clos.

Let $\rPv$ and $\rPs$ be the rate for processing a protocol $\calP$'s
message at the validator and switch, respectively.
(We assume the topology $\calT$ does not affect $\rPv$ and $\rPs$.)
Let $q$ denote the probability of a full round change 
when the leader is nonfaulty.

\subsubsection{Consensus Time Model}
\phantom{.}\\
$E\TPT$ depends critically on $q$.
Round change scenarios can be complex, 
and can vary greatly with how $\calP$ handles timeouts.
A full round change prompts a leadership change,
and can be caused by a faulty leader, or timer expiry at non-leaders.
We consider 3 possible events:
$\calE_1$ where the first leader for the consensus instance is faulty;
$\calE_2$ where the first leader is nonfaulty, but times out; and 
$\calE_3$ where the first leader is nonfaulty and does not time out.
Thus, 
\begin{equation}
	E(\TPT)=E(\TPT|\calE_1)r+ E(\TPT|\calE_2)(1-r)q
	+ E(\TPT|\calE_3)(1-r)(1-q),
\quad{\rm where\ } r=n_f/n. 
\label{eqn:key.eqn}
\end{equation}
It remains to estimate $E(\TPT|\calE_j)$ for $j=1,2,3$.
Let $\TPTTT=E(\TPT|\calE_3)$; 
this is the expected time if consensus finishes in 1 round 
(with no failures and timer expiry).

Throughout, we use ``first-order approximations''
that evaluate the most probable scenarios,
and ignore less likely, complicated possibilities.
In particular, we consider only a small number of faults.

\subsection{Simulator}
\label{sec:simulator}

We use a simulator (see Fig.~\ref{fig:architecture} in the appendix) 
to validate our performance model.
Message service times at validators and switches are exponentially distributed
by default, but these can be changed.
Service rate $\rPs$ is relative to $\rPv$,
and we set $\rPv=1/3$ arbitrarily.
Time unit for $\TPT$ in all our plots follows that for $1/\rPv$.
To factor out the impact of routing, 
the simulator uses ROMM routing~\cite{ROMMrouting.book2011},
where a random next hop is chosen from alternative paths with minimal hop count.

We do just 1 simulation run per data point,
so there are no confidence intervals for rigorously evaluating model accuracy.
(The consensus time is averaged over 500--2000 consensus instances per run.)
Our objective is to derive equations that provide performance insight,
and we are ready to sacrifice numerical accuracy in exchange.
The role of the simulator is more of a sanity check,
to provide an assurance that the assumptions and approximations 
do not take the model far from the simulated reality,
possibly making the extracted insights spurious.

\section{Clique Topology}
\label{sec:clique}

In this section, we factor out the topology by assuming the network is a clique.
The results also apply for a non-clique with switches 
that are fast enough that inter-validator message delays are negligible.
We divide the performance analysis into two cases: 
with/without full round change.

\subsection{No Full Round Change}
\label{sec:nofrc}

For now, assume initial timer value $\btl$ is large and there are 0 faults.
so there are no full round changes.
Thus, $r=0$ and $q=0$, and $E\TPT=E(\TPT|\calE_3)$
from Eqn.(\ref{eqn:key.eqn}),
so $E\TPT=\TPTTT$.

For HotStuff, the leader is a performance bottleneck 
(see Fig.~\ref{fig:MsgPattern}(b)).
It has to process $n$ votes (as leader) 
plus an additional message (as validator)
for each of the first 3 phases.
In the final phase, it need only process the first $n-f$ votes
(before broadcasting \decide) plus one message (like any other validator).
The total is $3(n+1)+(n-f)+1=4n-f+4$.
Thus, 
\begin{equation}
	{\rm expected\ HotStuff\ consensus\ time\ is\quad }
\THCCC=\frac{4n-f+4}{\rHv}=\frac{11n+13}{3\rHv} \quad{\rm if\ } n=3f+1.
\label{eqn:ETHC0}
\end{equation}
Note that this bottleneck analysis assumes the message queue is always nonempty,
so there is no need for queueing theory~\cite{Kleinrock}.

For IBFT, all validators have to process 
1 \preprepare, $n$ \prepare\ plus $n$ \commit\ messages, so
\begin{equation}
	{\rm expected\ IBFT\ consensus\ time\ is\quad }
\TICCC=\frac{2n+1}{\rIv} .
\label{eqn:ETIC0}
\end{equation}
\noindent
\underline{\it Model Validation and Comparison (clique; no full round change)}
\phantom{.}\\
The message processing times are random variables,
so their arrival and departure times are random.
Add to this the phase overlap among validators and between consecutive blocks,
and it is not obvious that the above simplistic derivations would work.
Nonetheless, we have validated Eqns.~(\ref{eqn:ETHC0}) and (\ref{eqn:ETIC0}) 
with the simulator (the straightline plots are omitted here,
in favor of other results).

HotStuff takes more phases to reach consensus than IBFT,
but we can see clearly from Eqns.~(\ref{eqn:ETHC0}) and (\ref{eqn:ETIC0})
that whether $\THCCC>\TICCC$ 
depends on the relative values of $\rHv$ and $\rIv$.
A similar comment applies if we compare how they scale with $n$,
since $\THCCC\approx\frac{11}{3\rHv}n$ and
$\TICCC\approx \frac{2}{\rIv} n$.

\subsection{With Full Round Change}
We can see from the simulation results in Fig.~\ref{fig:IntroSim}(a) that,
if full round change can happen --- because the timer value is too small
or there are faults --- then consensus time becomes non-monotonic.

\subsubsection{HotStuff (clique; with full round change)}
\label{sec:HSclique}
\phantom{.}\\ 
HotStuff has 4 phases, each with the leader broadcasting a message,
and each validator processing it and replying with a vote.
For the 4th phase, only $n-f$ votes are required before the leader broadcasts
the \decide\ messages.
The number of messages the leader needs to process is thus
\begin{equation}
	\mH=3(n_w+1)+n-f+1=3(n-n_f+1)+n-f+1=4n-3n_f-f+4\quad{\rm so\ }
	\THCCC =\frac{\mH}{\rHv} .
	\label{eqn:mH}
\end{equation}
A full round change occurs if the leader does not process $\mH $ messages
within $\btl$.

We now estimate $E(\THC|\calE_j)$ in Eqn.~(\ref{eqn:key.eqn}).
For $\calE_1$, the first leader is faulty and the number of subsequent rounds
is geometrically distributed.
The timer has exponential backoff, 
and the final consensus round has time $\THCCC$,
so $E(\THC|\calE_1)\approx \sum_{j=1}^\infty r^{j-1}(2^{j-1}\btl)
+ \THCCC = \THCCC+\frac{1}{1-2r}\btl$.

For $\calE_2$, the nonfaulty first leader times out.
This is similar to having a faulty leader, so $E(\THC|\calE_2)=E(\THC|\calE_1)$.
(We consider negligible the probability that both leaders of two consecutive rounds
time out.)
Thus Eqn.(\ref{eqn:key.eqn}) gives
\begin{equation}
E\THC=(\THCCC+\frac{1}{1-2r}\btl)r +(\THCCC+\frac{1}{1-2r}\btl)(1-r)q
	+\THCCC(1-r)(1-q)
	=\THCCC+\frac{r+(1-r)q}{1-2r}\btl .
	\label{eqn:ETHC}
\end{equation}

To estimate $q$, we consider the probability distribution for processing
$\mH$ messages.
If the processing time $X_i$ for the $i$th message has mean 
$\frac{1}{\rHv}$ and variance $\sigma^2$,
then we can use the Central Limit Theorem to approximate the processing time
$Y=\sum_{i=1}^{\mH}X_i \sim{\mathcal{N}}(\mH\frac{1}{\rHv},\mH\sigma^2)$
and $q=\Pr(Y>\btl)$.
\\

\begin{figure}
        \centering
        \begin{tabular}{cc}
                \includegraphics[width=0.45\textwidth]{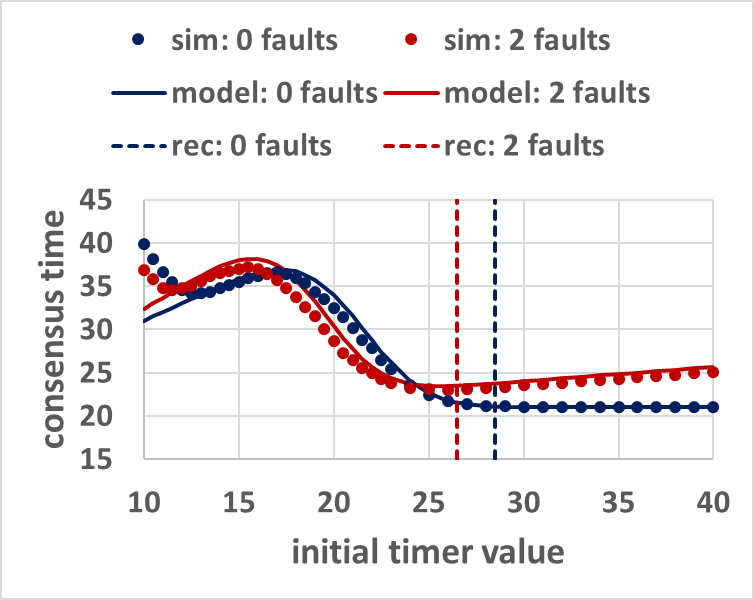} &
                \includegraphics[width=0.45\textwidth]{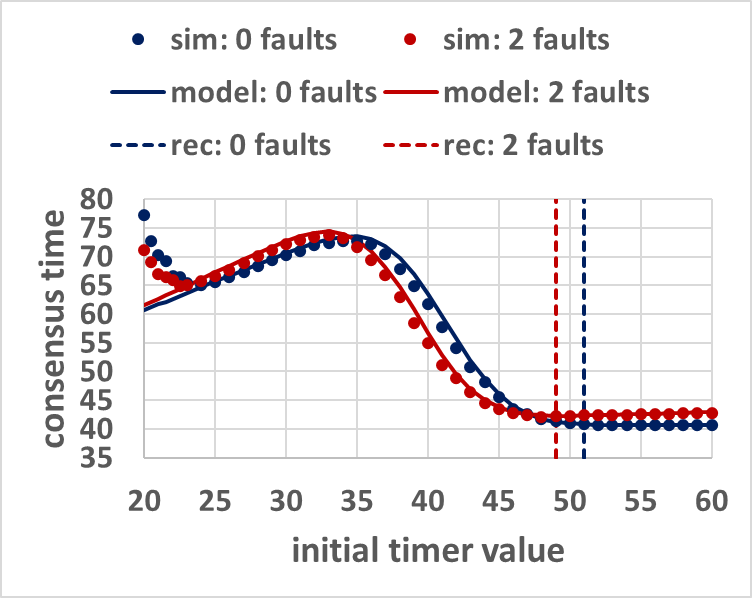} \\
		(a) HotStuff (clique, $n=16, n_f=0,2$)  & (b) HotStuff (clique, $n=32, n_f=0,2$)  \\
        \end{tabular}
	\caption{Comparison of Eqn.(\ref{eqn:ETHC}) 
	with simulation (with full round change).
	Vertical lines indicate $\btl^\ast$ (Eqn.(\ref{eqn:THCrec})).
	The unit of time follows that for $1/\rPv$ (see Sec.~\ref{sec:simulator}).}
	\label{fig:validTHC}
\end{figure}

\noindent
\underline{\it HotStuff Model Validation and Analysis (clique; with full round change)}
\phantom{.}\\
Fig.~\ref{fig:validTHC} compares the model (Eqn.~\ref{eqn:ETHC})
to simulated consensus time for $n=16$ and 32, with full round change.
It shows good agreement between model and simulator,
except for small $\btl$, where the divergence is expected
since ``second-order effects'' from complicated scenarios caused by
a too-small $\btl$ are no longer negligible.

We see from Eqn.(\ref{eqn:ETHC}) and the plot that, 
even when there are 0 faults ($r=0$),
$E\THC=\THCCC+q\btl$ increases from its minimum 
as $\btl$ decreases on the left, raising timeout probability.
If $\btl$ is sufficiently large, so nonfaulty leaders do not timeout,
then $q\approx 0$ and $E\THC\approx \THCCC+\frac{r}{1-2r}\btl$.
Thus, the tail on the right in Fig.~\ref{fig:validTHC} is linear in $\btl$,
and has gradient $\frac{r}{1-2r}$ that increases with number of faults $n_f$.
Note that this effect depends on $n_f$ and $n$ only through the ratio 
$r=\frac{n_f}{n}$.

How should $\btl$ be set?
It should be larger than $\THCCC$,
to provide some slack to accommodate the variance in $\THC$.
One possibility is to give an allowance of 3 standard deviations, i.e.
\begin{equation}
	\btl\leftarrow\btl^\ast=\mH\frac{1}{\rHv}
	+3\sqrt{\mH}\ \sigma \quad{\rm where\ } \mH=4n-3n_f -f+4.
	\label{eqn:THCrec}
\end{equation}
Note that $\rHv$ and $\sigma$ can be measured with one-off profiling of 
message processing time.

Fig.~\ref{fig:validTHC} shows this recommended $\btl^\ast$ gives a good
estimate of $\btl$ where $E\THC$ is minimum.

\subsubsection{IBFT (clique; with full round change)}
\label{sec:IBFTcliquefaults}
\phantom{.}\\
With $n_f$ crash faults, Eqn.(\ref{eqn:ETIC0}) becomes 
$\TICCC=\frac{2n_w +1}{\rIv}$.
The specified order for leaders in IBFT makes it harder to derive
closed-form expressions, so our model randomly chooses the leaders instead;
this simplification should work well for large $n$. 
Thus, like HotStuff, the event $\calE_1$ that $i$ validators in a row 
are faulty leaders has probability $r^i=(\frac{n_f}{n})^i$, and
$
E(\TIC|\calE_1)\approx \sum_{j=1}^\infty r^{j-1}(2^{j-1}\btl)
+ \TICCC +\frac{n_w}{\rIv}= \TICCC+\frac{1}{1-2r}\btl+\frac{n_w}{\rIv} ,
$
since there are $n_w$ round change messages.
For $\calE_2$, the first leader times out like a faulty leader.
There are possibly $n_w$ excess messages from the penultimate round,
and we estimate the final leader as processing $n_w(1-r)$ of them,
after factoring out the crashed validators.
Therefore,
$
	E(T^\calI_C|\calE_2)\approx E(\TIC|\calE_1) +\frac{n_w(1-r)}{\rIv}
	= \TICCC+\frac{1}{1-2r}\btl+\frac{n_w(2-r)}{\rIv} 
	$
so Eqn.(\ref{eqn:key.eqn}) gives
\begin{equation}
	E\TIC
=\TICCC+\frac{r+(1-r)q}{1-2r}\btl +\frac{r+(2-r)(1-r)q}{\rIv}n_w.
	\label{eqn:ETIC}
\end{equation}
Compared to $E\THC$ in Eqn.(\ref{eqn:ETHC}),
the third term here is for round change messages.

For HotStuff, we estimate $q$ by considering the messages a leader has to 
process before a timeout.
In IBFT, however, everyone sends messages to everyone, 
and a validator decides once it receives a quorum of \commit\ messages.
Thus, for a full round change to occur,
more than $f$ validators must undergo round change
before they broadcast their \commit\ messages.

Simulations show that a full round change is usually caused
by a leader who is delayed by a round change for the previous block.
To model this event, suppose the full round change occurs for consensus instance $i$ 
and, for consensus instance $i-2$, 
the $n_w$ validators were synchronized at the start of their \prepare\ phase.
A validator will need to process $\mI = n_w + n-f$ messages to complete
their \prepare\ and \commit\ phases before the second timer 
in Fig.~\ref{fig:MsgPattern}(a) expires.

Let $S_{i,k}$ be the time taken to process $\mI$ messages by validator $k$ in 
consensus instance $i$, and $X_k=S_{i-2,k} - \frac{1}{n_w}\sum_{j=1}^{n_w} S_{i-2,j}$.
Then $X_k$ measures how much faster or slower validator $k$ is relative
to the others at the end of instance $i-2$.
Let $\sigma^2$ be the variance in message processing time.
We then approximate the $X_k$ distribution by
$X_k \sim {\calN}(0,\mI(1+\frac{1}{n_w})\sigma^2)$.

When entering consensus instance $i-1$, faster validators (with $X_k<0$)
will enter the \prepare\ phase $-X_k$ time earlier and wait for the $\mI$
messages of slower validators.
The consensus instance $i$ round 1 leader, say validator $c$, 
is faster by $-X_c$, so the time it takes to process $\mI$ messages is
$Y=-X_c+S_{i-1,c}\sim\mathcal{N}(\mI\frac{1}{\rIv},
\mI(2+\frac{1}{n_w})\sigma^2)$ approximately,
and we estimate $q\approx\Pr(Y>\btl)$ with this distribution.

\begin{figure}
        \centering
        \begin{tabular}{cc}
                \includegraphics[width=0.45\textwidth]{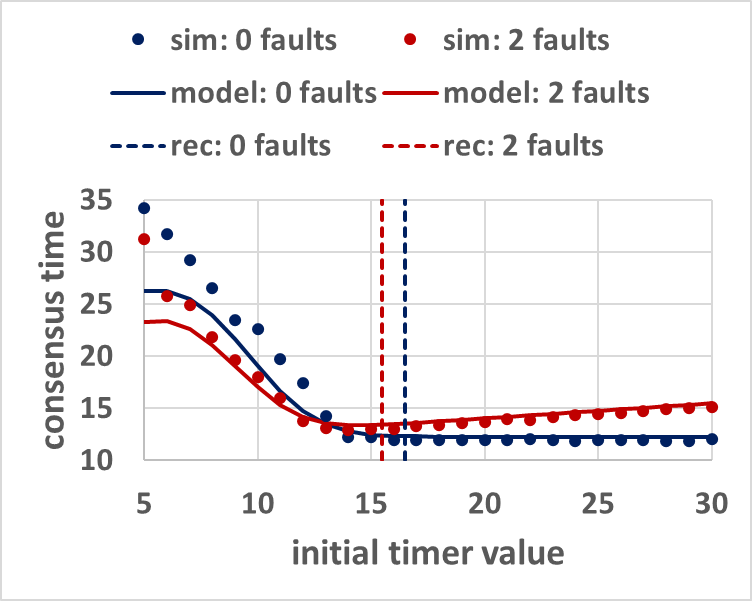} &
                \includegraphics[width=0.45\textwidth]{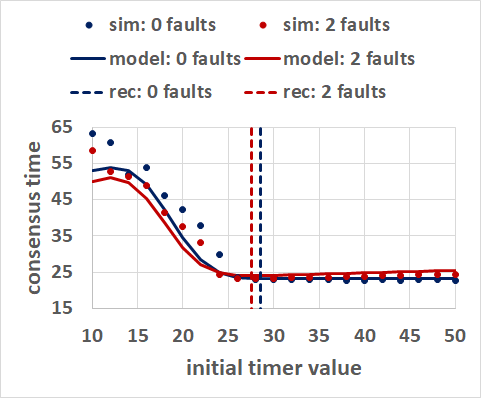} \\
		(a) IBFT (clique, $n=16$, $n_f=0,2$) & 
		(b) IBFT (clique, $n=32$, $n_f=0,2$) \\
        \end{tabular}
	\caption{Comparison of Eqn.(\ref{eqn:ETIC}) with simulation
	(with full round change).
	Leaders have a specified order in the simulation, 
	but are randomly selected in the model.
	Vertical dashed lines indicate $\tau_0^\ast$ (Eqn.(\ref{eqn:TICrec})).}
\label{fig:validTIC}
\end{figure}

\noindent
\underline{\it IBFT Model Validation and Analysis (clique; with full round change)}
\phantom{.}\\
Fig.~\ref{fig:validTIC} compares the IBFT model Eqn.(\ref{eqn:ETIC})
to simulated consensus time for $n=16$ and $n=32$.
The agreement in the tail for large $\btl$ is good, 
but diverges as $\btl$ gets smaller.
This is expected, since we needed multiple approximations to estimate $q$.

The tail (where $q\approx 0$) in Fig.~\ref{fig:validTIC} has the form
$E\TIC=\TICCC+\frac{r}{1-2r}\btl + \frac{r}{\rIv}(n-n_f)$;
for fixed $n_f$, this is still linear in $\btl$, 
with a gradient $\frac{r}{1-2r}$ (same as HotStuff) that increases with $n_f$
and depends on $n_f$ and $n$ through $r$.

How should the timer be initialized?
Using the same rule of thumb as before, it would be
\begin{equation}
	\btl\leftarrow\btl^\ast=\mI\frac{1}{\rIv}
	+3\sqrt{\mI(2+\frac{1}{n-n_f})}\ \sigma ,\quad{\rm where\ }
	\mI=2n-n_f-f.
	\label{eqn:TICrec}
\end{equation}
Fig.~\ref{fig:validTIC} shows this $\btl^\ast$ gives a good 
estimation of where $E\TIC$ is minimum.
Note that $\btl^\ast$ for IBFT is less sensitive to $n$ and $n_f$ for IBFT
than for HotStuff ($2n-f$ vs $4n-f$ and $-n_f$ vs $-3n_f$).

\section{Non-clique Topology}
\label{sec:nonclique}

Fig.~\ref{fig:IntroSim}(b) and (c) show that, for the Dragonfly, 
whether HotStuff is faster than IBFT depends on $n$ and the switch rate.
This is similarly true for Folded-Clos.
In this section, we model how the topology impacts consensus time $E\TPT$.
The approach is to focus on the validator and switch bottlenecks.

\subsection{No Full Round Change}
\label{sec:nofullrc}

As before, consider the case where $\btl$ is large
and validators are nonfaulty, so there is no full round change and $E\TPT=\TPTTT$.
Instead, we analyze the effect of changing $n$ and switch rate $r^\calP_s$.

\subsubsection{HotStuff (non-clique; no full round change)}
\label{sec:HSnorc}
\begin{figure}
\includegraphics[width=0.85\textwidth]{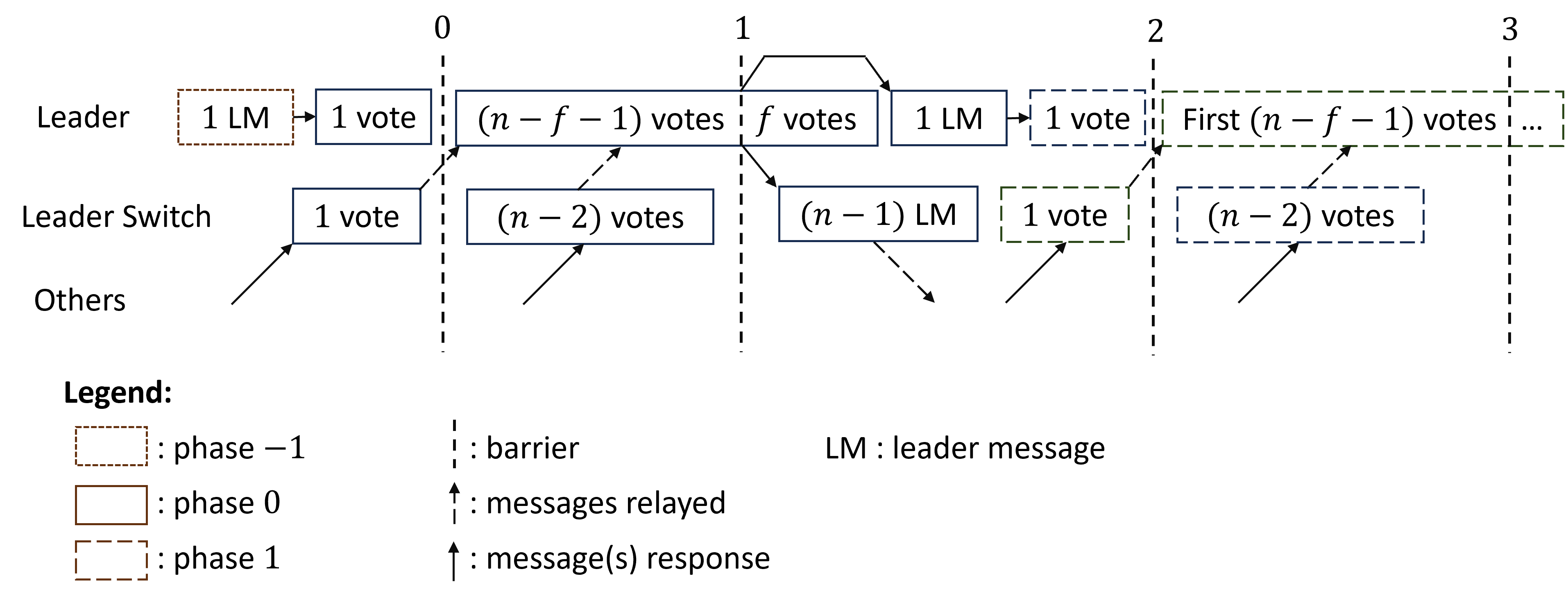}
\caption{
Progress in HotStuff.
	Barrier0: leader ready to process phase 0 (\prepare) votes
	from nonleaders.
	Barrier1: leader switch starts relaying leader's broadcast.
	Barrier2: leader ready to process phase 1 (\precommit) votes 
	from nonleaders.  
	\precommit, \commit\ and \decide\  phases have similar barriers.
	}
\label{fig:barrier}
\end{figure}

\phantom{.}\\ 
The HotStuff consensus time is mostly determined by the computation 
bottleneck at the leader and the communication bottleneck at the leader switch
(where the leader is).
Our analysis is based on the concept of {\it barriers}
(see Fig.~\ref{fig:barrier}).
For a specific round, we can take \prepare\ to be phase 0, 
\precommit\ to be phase 1, etc. and 
\decide\ of the previous consensus instance to be phase $-1$.

Barrier0 marks the time after a leader processes the leader message for the
previous consensus instance and its own vote;
this is when the leader is ready to process phase 1 votes from nonleaders.
Thus, the leader is guaranteed to have at least 1 of $n-f$ 
votes to process before the barrier.

Barrier1 marks the time after the leader 
has processed $n-f-1$ additional votes and sent a broadcast, and 
the leader switch has cleared its queue (of $n-2$ votes);
this is when the leader switch is ready to relay the $n-1$ broadcast messages.

Barrier2 marks the time after the leader switch has sent out the broadcast
and relayed the first incoming vote to the leader ($n-1+1$ messages),
and the leader has processed all $f$ remaining votes from phase 0,
plus 1 vote and its own leader message;
this is when the waiting pattern repeats and Barrier2 acts like Barrier0.

Similarly, Barrier3 is like Barrier1, etc.
This pattern repeats up till the \decide\ phase, 
where there is only the portion between Barrier0 and Barrier1.
After that, the remaining delay is for a validator to receive its \decide\
leader message to complete the consensus instance.

The time between Barrier0 and Barrier1 for each of HotStuff's 4 phases is
$\max\{\frac{n-f-1}{\rHv},\frac{n-2}{\rHs}\}$.
The time between Barrier1 and Barrier2 for each of the 3 phases before \decide\
is $\max\{\frac{f+2}{\rHv},\frac{n}{\rHs}\}$.
It takes time $\frac{\hT}{\rHs}+\frac{1}{\rHv}$ for a \decide\ message 
to reach a validator and be processed,
and another $\frac{\hT}{\rHs}+\frac{1}{\rHv}$ for the next leader to receive
and process a \newview\ message.
Thus,
\begin{equation} 
E\THT=\THTTT=4\max\left\{\frac{n-f-1}{\rHv},\frac{n-2}{\rHs}\right\}
                + 3\max\left\{\frac{f+2}{\rHv}, \frac{n}{\rHs}\right\}
		+2\big(\frac{1}{\rHv} + \frac{h_\calT}{\rHs}\big).
	\label{eqn:ETHT}
\end{equation}

\begin{figure}
        \centering
        \begin{tabular}{cc}
                \includegraphics[width=0.45\textwidth]{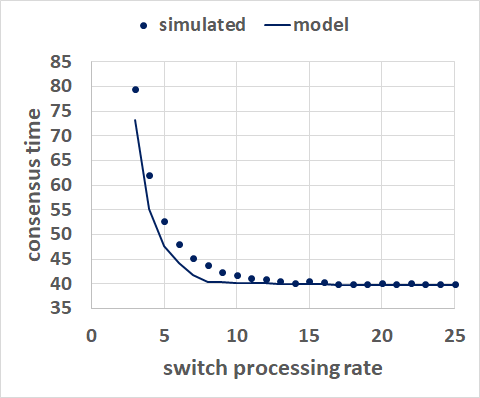} &
                \includegraphics[width=0.45\textwidth]{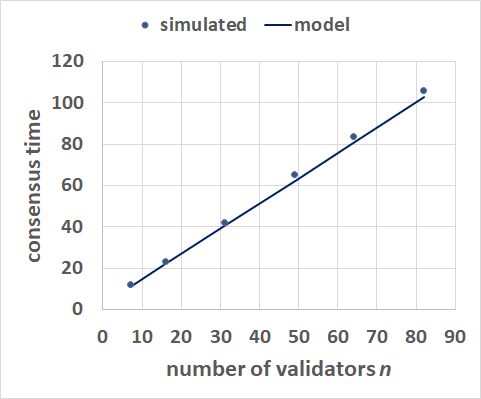} \\
		(a) Folded-Clos(8,4), $n=31$. & (b) Folded-Clos(10,5), $\rHs=9$. \\
        \end{tabular}
	\caption{Comparison of Eqn.(\ref{eqn:ETHT}) to HotStuff simulation
	(no full round change) for changes in (a) $\rHs$ and (b) $n$.}
	\label{fig:validTHF}
\end{figure}

\noindent
\underline{\it HotStuff Model Validation and Analysis 
(non-clique; no full round change)}
\phantom{.}\\
Fig.~\ref{fig:validTHF} compares Eqn.(\ref{eqn:ETHT}) to simulated HotStuff
consensus time for the Folded-Clos.
(The Dragonfly results are similar; this is not the case for IBFT.)
In the Appendix, we show how the model's accuracy 
for Fig.~\ref{fig:validTHF}(a) can be improved. 
Although that approximation is more accurate (see Fig.~\ref{fig:ETHFimproved}),
extracting performance insight is harder,
so we will continue with Eqn.(\ref{eqn:ETHT}).

We can tell from Eqn.(\ref{eqn:ETHT}) whether computation or communication 
dominates consensus time.
Switching dominates the first max if
$\frac{n-f-1}{\rHv}<\frac{n-2}{\rHs}$, 
i.e. $\rHs<\frac{3f+1-2}{3f-f-1}\rHv =\frac{3f-1}{2f}\rHv$,
and dominates the second max if 
$\frac{f+2}{\rHv}<\frac{n}{\rHs}$, i.e. $\rHs<\frac{3f+1}{f+2}\rHv$.
Thus, HotStuff consensus time is mostly caused by network delays if
$\rHs<\frac{3}{2}\rHv$ (independent of $n$).
This can happen if switch rate $\rHs$ is reduced by cross traffic
from other datacenter workloads (see Sec.~\ref{sec:key.eqn} and
Sec.~\ref{sec:multichain}).

Note that topologies with $\frac{\hT}{\rHs}\ll \frac{n}{\rHs}$
would have similar HotStuff performance.

\subsubsection{IBFT (non-clique; no full round change)}
\label{sec:IBFTtopo-nofrc}
\phantom{.}\\ 
Every validator in IBFT has to broadcast messages,
so the behavior is too complicated for a HotStuff-like 
leader-based barrier analysis.
We therefore simplify the analysis further to just comparing the computation
and communication bottlenecks,
so $\TITTT\approx\max\{\frac{m_v}{\rIv},\frac{\mIT}{\rIs}\}$,
where the leader must process $m_v=2n+1$ messages 
(see Fig.~\ref{fig:MsgPattern}(a)),
and $\mIT$ is the number of messages the leader switch in topology $\calT$ 
must relay.
There are 2 broadcasts and $(n-1)$ \preprepare\ messages,
so $\mIT=2\nIT+(n-1)$,
where each broadcast requires relaying $\nIT$ messages.

For Folded-Clos, each validator sends and receives $(n-1)+(n-1)$ messages
per broadcast.
An edge switch has $\ k_\calF(=\frac{n}{\nu_e}$) validators,
so (after subtracting double-counting) it relays
$\nIF=k_\calF 2(n-1)-k_\calF (k_\calF -1)$ messages.

For Dragonfly, a switch in any group $G$ has a link to 
another switch in some other group $G^\prime$, 
so it must relay $k_\calD (\nd-1)$ messages from the other validators in $G$ 
to the $k_\calD \nd$ validators in $G^\prime$, and vice versa, 
for each of the broadcasts.
Therefore,

\begin{align}
E\TIT=\TITTT &= \max\left\{\frac{2n+1}{\rIv},\frac{\mIT}{\rIs}\right\},
\ {\rm where\ } \mIF = 2(k_\calF 2(n-1) -k_\calF(k_\calF -1)) + (n-1),
	\ k_\calF =\frac{3n}{\ns}; \nonumber \\
	\ {\rm and\ } \mID &= 2(k_\calD 2(n-1) -k_\calD(k_\calD -1)) 
	+2k^2_\calD\nd(\nd-1)+ (n-1),
	\ k_\calD =\frac{n}{\ns}.
	\label{eqn:ETIT}
\end{align}

\begin{figure}
        \centering
	\begin{tabular}{cc}
                \includegraphics[width=0.45\textwidth]{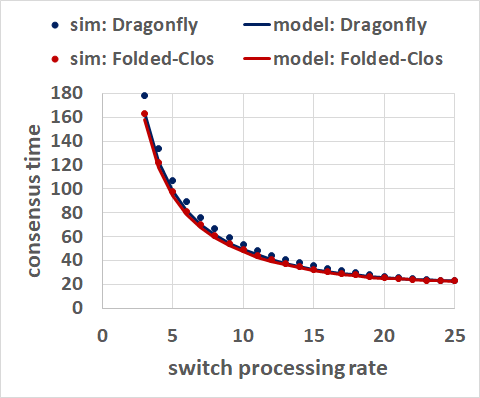} &
                \includegraphics[width=0.45\textwidth]{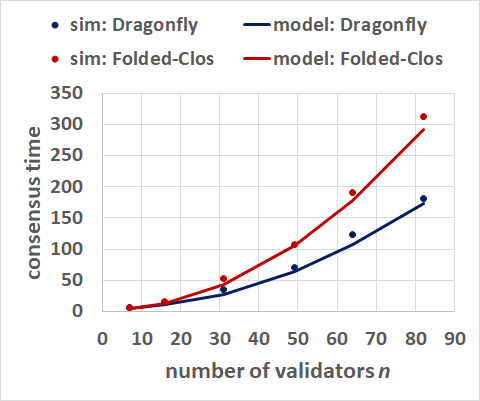} \\
		(a) $n=31$ & (b) $\rIs=9$ \\ 
        \end{tabular}
	\caption{Comparison of IBFT simulation (no full round change)
	to Eqn.(\ref{eqn:ETIT}) on Folded-Clos and Dragonfly
	for changing (a) switch rate $\rIs$ and (b) number of validators $n$.
	}
	\label{fig:validTIT}
\end{figure}

\noindent
\underline{\it IBFT Model Validation and Analysis (non-clique; no full round change)}
\phantom{.}\\
Fig.~\ref{fig:validTIT} shows Eqns.(\ref{eqn:ETIT}) 
gives a good fit for simulated consensus times for both topologies.
We now use the equations to compare HotStuff with IBFT,
and Dragonfly with Folded-Clos.
\\

\noindent
\underline{HotSutff vs IBFT (non-clique; no full round change)}
\label{sec:HSvIBFT}
\phantom{.}\\
If $\rPs$ has greater impact than $\rPv$,
then we can simplify Eqn.(\ref{eqn:ETHT}) for HotStuff to
$ E\THT=4\frac{n-2}{\rHs} + 3\frac{n}{\rHs} +\frac{2\hT}{\rHs}$,
which is linear in $n$.
If the switches dominate IBFT performance,
then Eqn.(\ref{eqn:ETIT}) gives
$E\TIT=\mIT/\rIs$ where $\mIF$ and $\mID$ are quadratic in $n$.
We can see this linear and quadratic behavior in Fig.~\ref{fig:IntroSim}(b).

However, suppose we fix $k_\calT$ 
and scale the number of switches by  $\ns=\frac{n}{k_\calT}$.
Then 
\begin{eqnarray}
	ET^\calI_\calF &=& (\frac{1}{\rIs})(
	6k_\calF(2(n-1) -(3k_\calF-1)) + (n-1)) \ {\rm for\ Folded-Clos}  \nonumber\\
	ET^\calI_\calD &=& (\frac{1}{\rIs})(
	2k_\calD (2(n-1)-(k_\calD -1)+2n\frac{\nd(\nd-1)}{\nd(\nd+1)})+(n-1))
	       \ {\rm for\ Dragonfly} \nonumber
        \label{eqn:ETIs}
\end{eqnarray}
which are linear in $n$ ($\frac{\nd(\nd-1)}{\nd(\nd+1)}\approx1$ for large $\nd$).

Although IBFT consensus time is quadratic in $n$, Fig.~\ref{fig:IntroSim}(b) 
shows that IBFT can be faster than HotStuff for small $n$;
Fig.~\ref{fig:IntroSim}(c) shows a similar performance crossover point
if the switch rate is sufficiently large.
We can locate these crossover points by analyzing the closed-form expressions.

From Eqn.(\ref{eqn:ETHT}), 
$E\THT = 4\frac{n-f-1}{\rHv}+3\frac{f+2}{\rHv}+\frac{2}{\rHv}
	=(4n-f+4)\frac{1}{\rHv}$,
while Eqn.(\ref{eqn:ETIT}) gives $E\TIT=\mIT/\rIs$.
The crossover $E\THT=E\TIT$ occurs at
\begin{equation}
	\rIs =\frac{\mIT}{4n-f+4}\rHv.
	\label{eqn:Xover}
\end{equation}

\begin{figure}
        \centering
        \begin{tabular}{cc}
\includegraphics[width=0.45\textwidth]{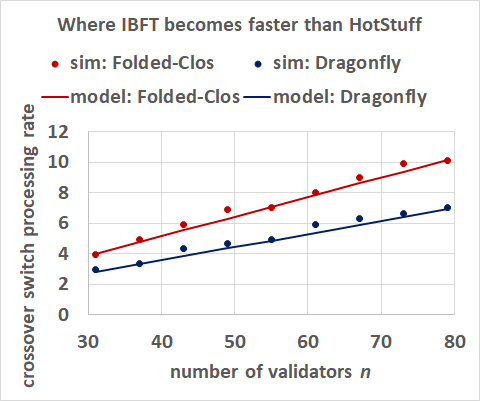} &
\includegraphics[width=0.45\textwidth]{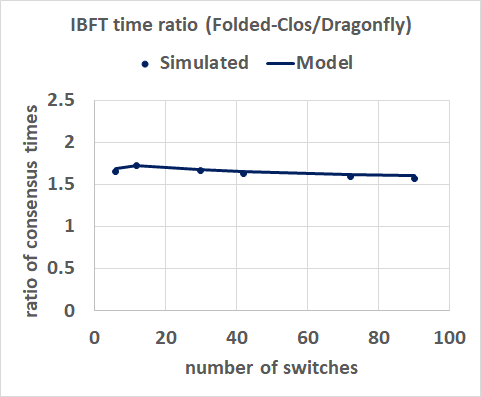} \\
		(a) HotStuff vs IBFT & (b) Dragonfly vs Folded-Clos \\
        \end{tabular}
	\caption{Comparing simulation (no full round change)
	         to (a) Eqn.(\ref{eqn:Xover}) for
	         switch processing rate $\rIs$ where 
	         IBFT becomes faster than HotStuff;
		 (b) Eqn.(\ref{eqn:TimeRatio}) for
		 IBFT consensus time ratio between 
		 Folded-Clos and Dragonfly for $k=2$ 
		 (the simulated values are for $\nd(\nd+1)$ divisible by 3). }
	\label{fig:Xover}
\end{figure}

Fig.~\ref{fig:Xover}(a) compares Eqn.(\ref{eqn:Xover}) to
the simulated HotStuff/IBFT crossover points for the Dragonfly and Folded-Clos.
The plot shows good agreement, 
despite many assumptions and approximations in the model.

One can do a similar analysis for a different choice of dominant terms in
Eqns.(\ref{eqn:ETHT}) and (\ref{eqn:ETIT}).
\\

\noindent
\underline{Dragonfly vs Folded-Clos (IBFT; no full round change)}
\phantom{.}\\
HotStuff's consensus times on Dragonfly and Folded-Clos 
are similar (see Eqn.(\ref{eqn:ETHT})).
This is also true for IBFT if computation time is the dominant factor,
and the $\max$ in Eqn.(\ref{eqn:ETIT}) resolves to $\frac{2n+1}{\rIv}$.

Consider now the case where IBFT performance is dominated by switching time.
Suppose both topologies have the same number of switches $\ns$,
$k_\calF=\frac{n}{\nue}=\frac{3n}{\ns}$
and $k_\calD=\frac{n}{\nue}=\frac{n}{\ns}$.
Then
\begin{equation}
	\frac{E\TIF}{E\TID} = \frac{\mIF}{\mID} 
	= \frac{2k_\calF(2(n-1) - (k_\calF -1))+(n-1)}
	       {2k_\calD(2(n-1) - (k_\calD -1) + 2k_\calD(\nd(\nd-1)))+(n-1)}
	\approx\frac{1.5}{1+\frac{1}{8k_\calD}}
	\label{eqn:TimeRatio}
\end{equation}
since $k_\calD\nd(\nd-1)=n\frac{\nd(\nd-1)}{\nd(\nd+1)}$
and $k_\calF=3k_\calD$.
Fig.~\ref{fig:Xover}(b) shows that, even for small $\nd$,
$\frac{E\TIF}{E\TID}$ is almost constant.
The insight here is that, for the same number of switches, 
Folder-Clos is consistantly slower than the Dragonfly
(which spreads the $n^2$ communication workload over more switches).

\subsection{With Full Round Change}

\subsubsection{HotStuff (non-clique; with full round change)}
\label{sec:HStopofaults}
\phantom{.}\\ 
Here, we revisit the barrier analysis in Sec.~\ref{sec:HSnorc}.  
For Barrier1, the leader will still need to process $n-f$ messages to 
broadcast the message for the next phase.
However, the leader switch will need to clear $n_w-2$ votes
before it starts relaying messages from the leader.
The time required is thus $\max\{\frac{n-f-1}{\rHv},\frac{n_w-2}{\rHs}\}$.
Similarly, for Barrier2, the leader will still broadcast $n+1$ messages,
but will only have $f-n_f +2$ messages (instead of $f+2$)
left to process since it received $n-n_f$ responses for that phase.
Thus, with $n_f\ne0$, Eqn.(\ref{eqn:ETHT}) becomes
\begin{equation}
E\THT=4\max\left\{\frac{n-f-1}{\rHv},\frac{n-n_f-2}{\rHs}\right\}
                + 3\max\left\{\frac{f-n_f+2}{\rHv}, \frac{n}{\rHs}\right\}
		+2(\frac{1}{\rHv}+\frac{\hT}{\rHs})
        +\frac{r+(1-r)q}{1-2r}\btl .
        \label{eqn:ETHTrc}
\end{equation}
(Note the impact of topology in $h_\calT$ is the same as Eqn.(\ref{eqn:ETHT}).)
We then estimate $q$ as $\Pr(Y>\btl)$, where $Y\sim\calN(\mu,\sigma^2)$, 
$\mu$ as in Eqn.(\ref{eqn:ETHTrc}) and 
\begin{equation}
	\sigma^2=4\max\{(n-f-1)\sigma_v^2, (n-n_f-2)\sigma_s^2\}
	+3\max\{(n-n_f+2)\sigma_v^2, n \sigma_s^2\} +2\sigma_v^2 + 2h_\calT \sigma_s^2
	\label{eqn:ETHTrc-sigma}
\end{equation}
where $\sigma_v$ and $\sigma_s$ are standard deviations for
validator and switch service times.
Here, we use crude approximations $Var(X+Y)\approx VarX + Var Y$ and
$Var\max\{X,Y\}\approx \max\{VarX,VarY\}$.

\begin{figure}
        \centering
        \begin{tabular}{ccc}
\includegraphics[width=0.31\textwidth]{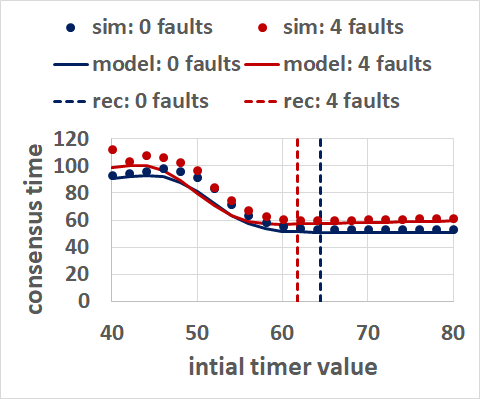} &
\includegraphics[width=0.31\textwidth]{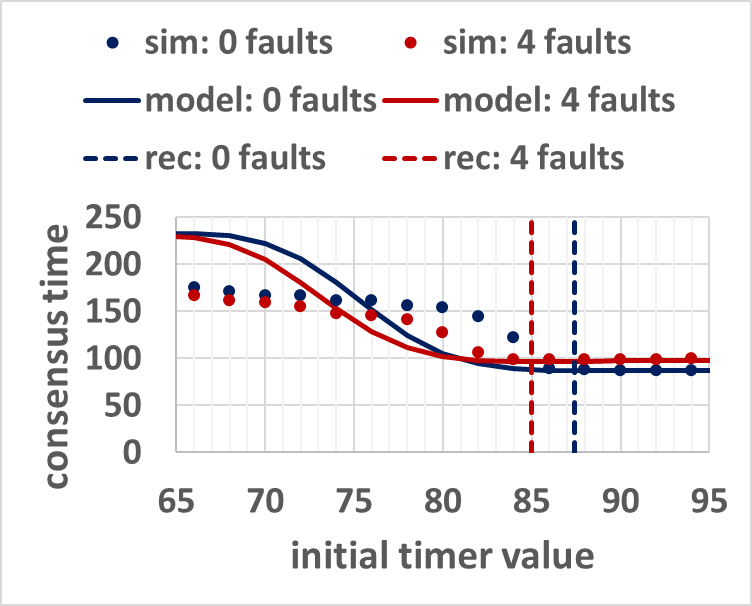}&
\includegraphics[width=0.31\textwidth]{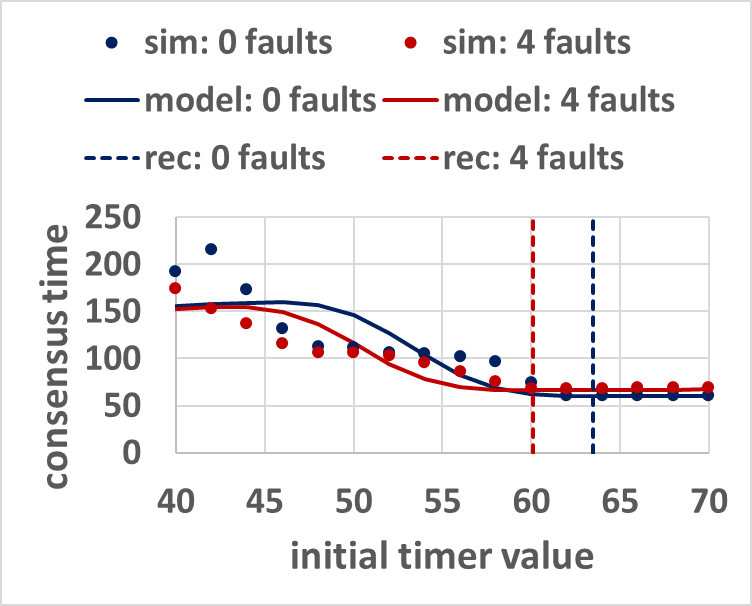} \\
		(a) Dragonfly(4) and $n=40$.
		& (b) Folded-Clos(8,4) and $n=40$.
		& (c) Dragonfly(4) and $n=40$. \\
        \end{tabular}
	\caption{Comparison of simulation to model for 
	(a) HotStuff (similar for Dragonfly and Folded-Clos)
	using Eqns.(\ref{eqn:ETHTrc}) and (\ref{eqn:ETHTrc-sigma}),
	(b) IBFT Eqn.(\ref{eqn:ETITfaults}) for Folded-Clos 
	and (c) IBFT Eqn.(\ref{eqn:ETITfaults}) for Dragonfly.}
        \label{fig:NoncliqueFaults}
\end{figure}

\subsubsection{IBFT (non-clique; with full round change)}
\label{sec:IBFTnonclique-frc}
\phantom{.}\\ 
For IBFT, we revisit the clique model in Sec.~\ref{sec:IBFTcliquefaults}
for full round change.
To factor in the topology, we consider the case where 
the switches are the bottleneck, so
$\TITTT=\frac{\mIT}{\rIs}=\frac{2\nIT+(n-1)}{\rIs}$ from Eqn.(\ref{eqn:ETIT}).
Each validator now sends $(n-1)$ messages and receives $(n_w-1)$ messages
per broadcast, so
$\nIF=k_\calF(n-1+n_w -1)-k_\calF(k_\calF-1)$ and
$\nID=k_\calD(n-1+n_w -1)-k_\calD(k_\calD-1) +2k^2_\calD\nd(\nd-1)(1-r)$
after factoring out crashed validators with $(1-r)$.
The $n_w$ messages in Eqn.(\ref{eqn:ETIC}) is now $\nIT$, so we get 
\begin{align}
	E\TIT &= \frac{2\nIT+(n-1)}{\rIs} +
        \frac{r+(1-r)q}{1-2r}\btl 
	+\frac{r+(2-r)(1-r)q}{\rIv}\nIT, 
	\ {\rm where\ } \nonumber\\
	\nIF &= k_\calF(n+n_w-k_\calF -1)
	\ {\rm and\ } \nID = k_\calD(n+n_w-k-1)+2k^2_\calD\nd(\nd-1)(1-r).
        \label{eqn:ETITfaults}
\end{align}
Like for Eqn.(\ref{eqn:ETIT}), Eqn.(\ref{eqn:ETITfaults}) is quadratic in $n$
for fixed $\ns$, but linear in $n$ for fixed $k_\calT$.

To estimate $q$ in Eqns.(\ref{eqn:ETITfaults}),
we combine the estimates in 
Sec.~\ref{sec:IBFTcliquefaults} (for full round change)
and Sec.~\ref{sec:IBFTtopo-nofrc} (for non-clique).
Per consensus instance, a validator only needs to process $m_v=\mIT-f+n_f -1$ 
or $m_s=\mIT-k(f-n_f)-(n-1)$ messages 
(see Eqn.(\ref{eqn:ETIT}) for $\mIT$)
in total before reaching consensus.
The consensus time model is then
$Y\sim\calN(\max\{\frac{m_v}{\rIv},\frac{m_s}{\rIs}\},
\max\{m_v(2+\frac{1}{n_w})\sigma^2_v,m_s(2+\frac{1}{n_w})\sigma^2_s\})$,
where $\sigma^2_v$ and $\sigma^2_s$ are variances of message processing times
at the validator and switch respectively.
As before, we use 3 standard deviations for recommended $\btl^\ast$.

\noindent
\underline{\it Model with full round change for non-clique topology:
Validation and Analysis}
\phantom{.}\\
For HotStuff, Fig.~\ref{fig:NoncliqueFaults}(a) shows 
Eqn.(\ref{eqn:ETHTrc}) and the recommended $\btl^\ast$
fit the simulation reasonably well, 
despite the rough approximation in Eqn.(\ref{eqn:ETHTrc-sigma}) for $q$.
Similar to the analysis of Eqn.(\ref{eqn:ETHT}),
$E\THT$ is mostly caused by network delays if
$\rHs<\frac{n-n_f-2}{n-f-1}\rHv=\frac{2f+(f-n_f)-1}{2f}\rHv$,
so $n_f\ne0$ lowers the threshold on $\frac{\rHs}{\rHv}$
from $\frac{3}{2}$ in Sec.~\ref{sec:HSnorc} to $1$.

For IBFT, however, Fig.~\ref{fig:NoncliqueFaults}(b) and (c) show
Eqn.(\ref{eqn:ETITfaults}) does not fit the simulation well.
Even so, the $\btl^\ast$ values are accurate,
which indicate the model's location of the curve is on target.

We therefore extend Sec.~\ref{sec:nofullrc}'s HotStuff vs IBFT analysis.
Their comparison depends on $q$ and $r$,
so we only consider large $\btl$, where $q\approx0$.
Again, we focus on the case where the protocol's performance is
determined by $\rHv$ and $\rIs$.
Then, Eqn.(\ref{eqn:ETHTrc}) and (\ref{eqn:ETITfaults}) give
$
E\THT=\frac{4(n-f-1)+3(f-n_f +2)+2}{\rHv}+\frac{r}{1-2r}\btl
=\frac{4n-f-3n_f +4}{\rHv} +\frac{r}{1-2r}\btl$ 
and
$E\TIT=\frac{2\nIT+(n-1)}{\rIs} + \frac{r}{1-2r}\btl +\frac{r}{\rIs}\nIT$.
(Note that, like for the clique in Sec.~\ref{sec:HSclique},
the tails $\frac{r}{1-2r}\btl$ are the same.) 
Thus the crossover $E\THT=E\TIT$ occurs at
$ \rIs=\frac{(2+r)\nIT+(n-1)}{4n-f-3n_f+4}\rHv$,
which generalizes Eqn.(\ref{eqn:Xover}) for $n_f\ne0$.

\section{Conclusions}
\label{sec:conclusions}

We conclude by highlighting some insights, derived from our analytical model
on BFT consensus time, that are hard to obtain from simulations:
how the initial timer value $\btl$ should be set,
and how it should scale differently with $n$ for HotStuff and IBFT
(Eqns.(\ref{eqn:THCrec}), (\ref{eqn:TICrec}));
how the two protocols have the same tail $\frac{r}{1-2r}\btl$,
where $r$ is the fraction of crash faults
(Eqns.(\ref{eqn:ETHTrc}), (\ref{eqn:ETITfaults}));
how the bottleneck for HotStuff depends on $\rHs/\rHv$
(Sec.~\ref{sec:HSnorc}, \ref{sec:HStopofaults});
how IBFT consensus time scales quadratically with $n$ for fixed number of
switches and linearly for fixed number of validators per edge switch
(Eqns.(\ref{eqn:ETIT}), (\ref{eqn:ETITfaults}));
how the performance crossover between HotStuff and IBFT depends on $\rIs/\rHv$
and scales with $n$ (Eqn.(\ref{eqn:Xover}), Fig.~\ref{fig:Xover}(a));
and how IBFT's consensus time ratio for Folded-Clos and Dragonfly is 
independent of $n$ and number of switches 
(Eqn.(\ref{eqn:TimeRatio}), Fig.~\ref{fig:Xover}(b)).
One can similarly draw more conclusions by considering a different choice
of cases (in the $\max$ terms and bottlenecks).

We believe our modeling approach for IBFT can be used for analyzing
other BFT protocols with similar message patterns,
such as PBFT and BFT-SMART~\cite{BFTsmartDSN2014}
(after including details like the reconfiguration protocol).
It can also be combined with the approach for HotStuff
to model a protocol like Zyzzyva~\cite{ZyzzyvaTOCS2009},
which has a HotStuff-like message pattern for a ``fast path'' 
and a IBFT-like message pattern for a ``slow path''.

Similarly, our approach to analyzing the impact of the Dragonfly 
and the Folded-Clos should be applicable to 
other direct networks like the Slim-Fly~\cite{SlimFlySC2014}
and indirect networks like the Hedera~\cite{HederaNSDI2010}. 

The analysis suggests some tips for analytical modeling that should apply 
to other consensus protocols:
choose a metric that focuses on the protocol (Sec.~\ref{sec:key.eqn});
use the switch rate to factor out other network traffic 
(Sec.~\ref{sec:HSnorcImproved});
use bottleneck analysis instead of queueing theory (Sec.~\ref{sec:nofrc});
use barrier analysis to mark the time when a node or switch changes phase
(Fig.~\ref{fig:barrier});
do not shy away from crude approximations (e.g. using $\max\{EX,EY\}$ 
instead of $E\max\{X,Y\}$) that provide analytically tractable closed-form
expressions --- they may suffice to model the first-order effects
and yield accurate conclusions (e.g. Fig.~\ref{fig:Xover}).

That said, it is clear that our modeling approach has reached its limit
when modeling IBFT for non-clique topologies.
Fig.~\ref{fig:NoncliqueFaults}(b) and (c) suggest that some other modeling
element is necessary for $q$ to shape the curve to fit the data.

\bibliographystyle{ACM-Reference-Format}
\bibliography{bc}

\appendix

\section{Appendix}

For easy reference, Table~\ref{tab:term.notation}
lists the frequently-used notation used in the paper.

\begin{figure}
        \centering
        \includegraphics[width=1.0\textwidth]{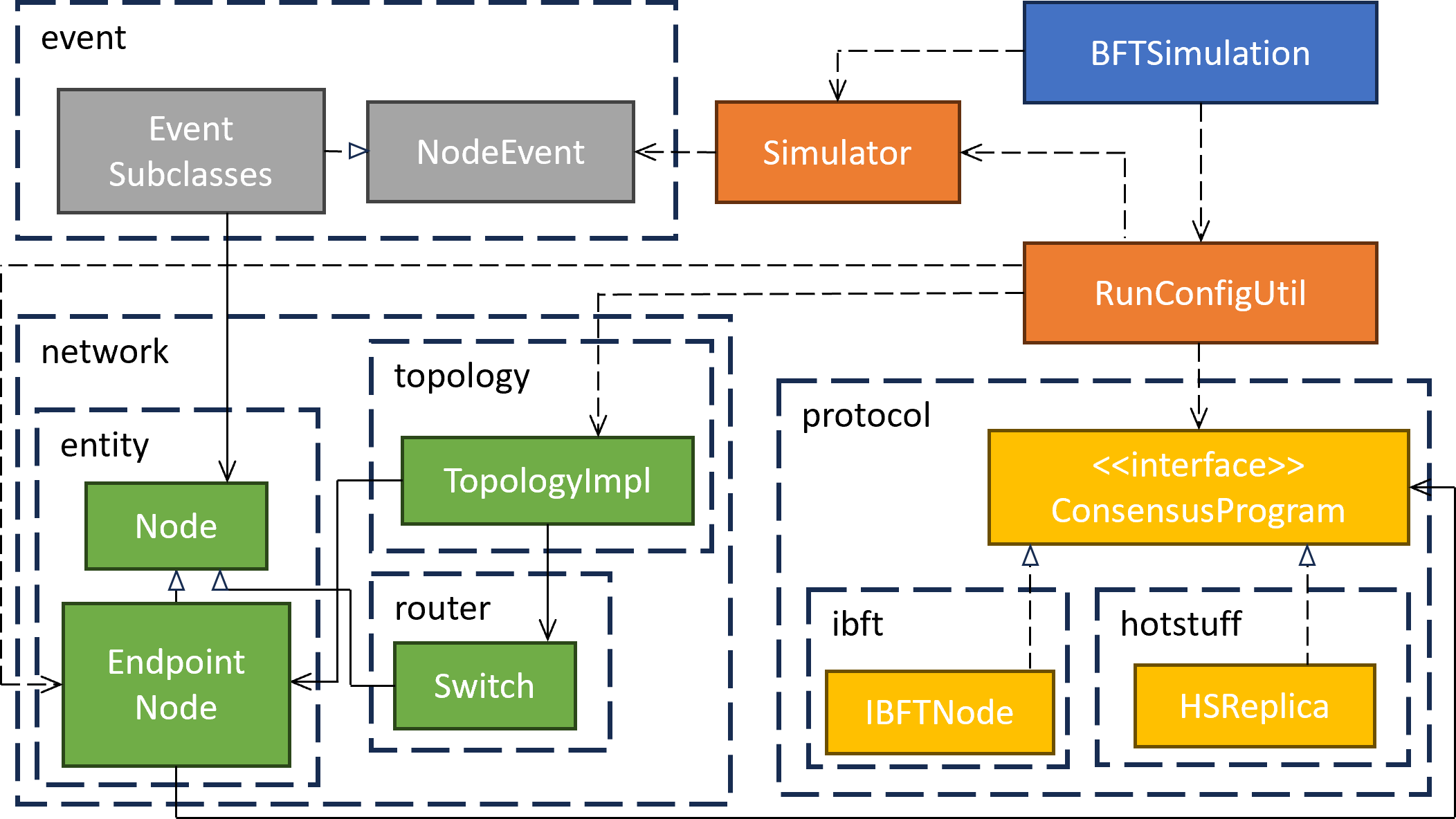}
        \caption{Architecture diagram of the main components of the simulator program. \label{fig:architecture}}
\end{figure}

The simulator we use for model validation is implemented in Java,
with some 3000 lines of code.
(The simulator is on {\tt GitHub}
and will be released with publication of this paper.)
It has protocol and network packages that can facilitate implementation
of different consensus protocols and network topologies.
Fig~\ref{fig:architecture} provides a high-level architecture diagram 
of the simulation program.

\begin{table}
\caption{Notation for parameters and variables.}
	\label{tab:term.notation}
	\begin{tabular}{|c|c|l|}
\hline
protocol & $\calP$ & consensus protocol $\calP=\calI$ for IBFT, 
	                      $\calP=\calH$ for HotStuff \\
	 & $\btl$  & initial value for timer \\
	 & $n$     & number of validators \\
	 & $n_w$   & number of working (i.e. nonfaulty) validators \\
	 & $n_f$   & number of faulty validators \\
	 & $r$     & fraction of faulty validators, $r=\frac{n_f}{n}$ \\
	 & $f$     & bound on $n_f$, $n_f\le f$, $n=3f+1$ \\
\hline
		topology & $\calT$ & network topology $\calT=\calC$ (Clique) or
			      $\calD$ (Dragonfly) or $\calF$ (Folded-Clos)\\
         & $\ns$   & number of switches \\
	 & $\nu_e$ & number of edge switches, 
		$\nu_e=\frac{\ns}{3}$ for Folded-Clos, $\nu_e=\ns$ for Dragonfly  \\
	 & $\nn$ & number of links to level 2 from a level 1 
		   switch in Folded-Clos \\
         & $\nd$ & number of links from a Dragonfly switch \\
         & $\ng$ & number of groups in a Dragonfly, $\ng=\nd+1$ \\
 & $k_\calT$  & number of validators per edge switch, $k_\calT=\frac{n}{\nu_e}$ \\
	 & $h_\calT$ & average number of network hops between two validators \\
\hline
performance & $\TPT$ & consensus time for protocol $\calP$ 
		       over topology $\calT$\\
       & $\rPv$ & service rate for a protocol $\calP$ message at a validator \\
       & $\rPs$ & service rate for a protocol $\calP$ message at a switch \\
& $m_\calT^\calP$ & number of $\calP$ messages a bottleneck switch 
		    in topology $\calT$ must process \\
		& $n_\calT^\calP$ & number of broadcast messages in $m_\calT^\calP$\\
            & $q$  & $q$ is probability Pr(full round change | nonfaulty leader)\\
	    & $\calE_1$ & event for faulty first leader \\
	    & $\calE_2$ & event for nonfaulty first leader, but times out \\
	    & $\calE_3$ & event for nonfaulty first leader, no timeout \\ 
	    & $\TPTTT$  & $E(\TPT|\calE_3)$ \\
	    & $\mathcal{N}(\mu_\ast,\sigma_\ast^2)$ & normal distribution 
		          with mean $\mu_\ast$ and variance $\sigma_\ast^2$ \\
\hline
\end{tabular}
\end{table}

\subsubsection{Competing network traffic}
\label{sec:multichain}
\phantom{.}\\ 
We model background and cross traffic by reducing the switch rate $\rPs$.
To validate this approach, we run $c=1,2,3$ chains concurrently
in the simulator, without changing the switch processing rate,
but with validator processing rate increased by $c$
(to model additional terminal nodes).
On the other hand, the model has switch processing rate $\rPs/c$,
and $\rPv$ unchanged.

\begin{figure}
        \centering
        \begin{tabular}{c}
		\includegraphics[width=0.5\textwidth]{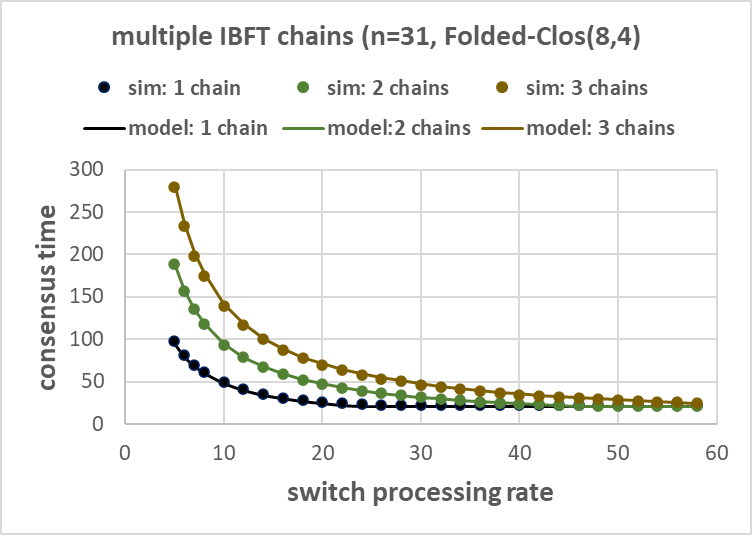}
        \end{tabular}
\caption{
Model with switch rate $\rIs/c$ fits consensus time for $c$ simulated chains 
for IBFT on a Folded-Clos$(8,4)$ with switch rate $\rIs$.
	The unit of time (on the vertical axis) 
	follows that for $1/\rPv$ (see Sec.~\ref{sec:simulator}).}
	\label{fig:multichain}
\end{figure}

Fig.~\ref{fig:multichain} shows a good fit between this model and the simulation
for IBFT on a Folded-Clos,
confirming our intuition that
we can model competing network traffic by having a single chain
but a slower switch rate.

\subsubsection{HotStuff (non-clique; no full round change; improved approximation)}
\label{sec:HSnorcImproved}
\phantom{.}\\ 
Fig.~\ref{fig:validTHF}(a) shows a significant difference between 
Eqn.(\ref{eqn:ETHT}) and simulated HotStuff consensus time.
We now show one way of improving the accuracy.

Let $\XHv$ and $\XHs$ be random variables for message processing times
at a validator and switch.
Then Eqn.(\ref{eqn:ETHT}) can be written as
\[
E\THT=4\max\{(n-f-1)E\XHv,(n-2)E\XHs\}
                + 3\max\{(f+2)E\XHv, nE\XHs\}
		+2E\XHv + {2h_\calT}E\XHs .
\]
Now, $(n-f-1)E\XHv=EX$ where $X$ is the sum of $n-f-1$ message processing
times at a validator.
Similarly, $(n-2)E\XHs=EY$ where $Y$ is the sum of $n-2$ message processing 
times at a switch.
Eqn.({\ref{eqn:ETHT}) thus uses $\max\{EX,EY\}$ to approximate $E\max\{X,Y\}$.
Using the Central Limit Theorem, we have
$X\sim\calN((n-f-1)E\XHv,(n-f-1)Var\XHv)$ and
$Y\sim\calN((n-2)E\XHs,(n-2)Var\XHs)$.
Therefore~\cite{max-gaussian},
$E\max\{X,Y\}\approx G(EX,VarX, EY,VarY)$ where 
\[
G(u_1,v_1,u_2,v_2)
=u_1\Phi\left(\frac{u_1-u_2}{\sqrt{v_1+v_2}}\right)
+u_2\Phi\left(\frac{u_2-u_1}{\sqrt{v_1+v_2}}\right)
+\sqrt{v_1+v_2}\phi\left(\frac{u_1-u_2}{\sqrt{v_1+v_2}}\right),
\]
where $\phi$ and $\Phi$ are the standard normal density and
cumulative distribution functions. 
We thus get the following alternative to Eqn.(\ref{eqn:ETHT}):
\begin{eqnarray}
	E\THT &=& 4G((n-f-1)\frac{1}{\rHv},(n-f-1)\sigma_v^2,
	(n-2)\frac{1}{\rHs}, (n-2)\sigma_s^2) \nonumber \\
	&\phantom{=}& + 3G((f+2)\frac{1}{\rHv}, (f+2)\sigma_v^2, 
	                    n\frac{1}{\rHs}, n\sigma_s^2) 
                +\frac{2}{\rHv}+\frac{2h_\calT}{\rHs}
\label{eqn:ETHFimproved}
\end{eqnarray}
where $\sigma_v^2=Var\XHv$ and $\sigma_s^2=Var\XHs$.

Fig.~\ref{fig:ETHFimproved} shows that Eqn.(\ref{eqn:ETHFimproved})
	gives a better approximation than Eqn.(\ref{eqn:ETHT}).
	However, this improved accuracy is at the expense of analytical tractability: 
	one cannot use Eqn.(\ref{eqn:ETHFimproved}) to compare HotStuff to IBFT,
	like we did with Eqn.(\ref{eqn:ETHT}) in Sec.~\ref{sec:nofullrc}.

\begin{figure}
        \centering
        \begin{tabular}{c}
                \includegraphics[width=0.5\textwidth]{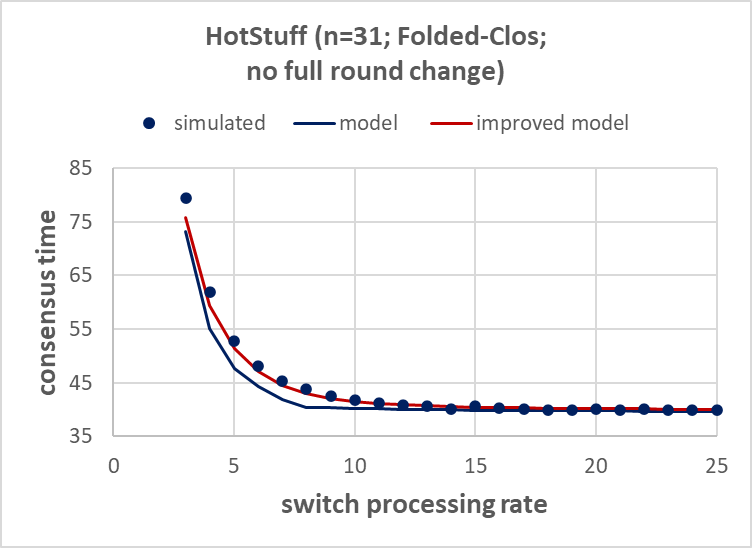} 
        \end{tabular}
	\caption{Fig.~\ref{fig:validTHF}(a) revisited: 
	Comparison of simple model Eqn.(\ref{eqn:ETHT})
	and improved model Eqn.(\ref{eqn:ETHFimproved}) to simulation for
	HotStuff on Folded-Clos(8,4).}
	\label{fig:ETHFimproved}
\end{figure}

\end{document}